\documentclass[%
 reprint,
 amsmath,amssymb,
prx,
floatfix,
superscriptaddress
]{revtex4-2}

\usepackage{mathtools}
\usepackage{graphicx}% Include figure files
\usepackage{dcolumn}% Align table columns on decimal point
\usepackage{bm}% bold math

\usepackage[caption=false]{subfig}

\usepackage{algorithm}
\usepackage[noend]{algpseudocode}
\algrenewcommand\algorithmicdo{}

\makeatletter
\renewcommand{\ALG@name}{Procedure}
\makeatother

\usepackage[linktocpage=true,
  colorlinks=true, 
  pdfborder={0 0 0},
  linkcolor=blue,
  citecolor=red,
  filecolor=yellow,
  urlcolor=blue,
  bookmarks,
  pdfauthor={},
]{hyperref}

\usepackage{orcidlink}

\usepackage{ifthen}
\newcounter{is_qcircuit_used}
\newcounter{are_figs_merged}
\newcommand{\argmin}{\mathop{\rm arg~min}\limits}

\begin{document}

\preprint{APS/123-QED}

\title{
Channel-agnostic finite-temperature phase estimation averaged over variable grids:\\ reconstruction of Green's function for dynamical mean-field theory 
}

\author{Taichi Kosugi\orcidlink{0000-0003-3379-3361}}
\email{kosugi.taichi@gmail.com}
\affiliation{
Quemix Inc.,
Taiyo Life Nihombashi Building,
2-11-2,
Nihombashi Chuo-ku, 
Tokyo 103-0027,
Japan
}

\affiliation{
Department of Physics,
The University of Tokyo,
Tokyo 113-0033,
Japan
}

\author{Hirofumi Nishi\orcidlink{0000-0001-5155-6605}}
\affiliation{
Quemix Inc.,
Taiyo Life Nihombashi Building,
2-11-2,
Nihombashi Chuo-ku, 
Tokyo 103-0027,
Japan
}

\affiliation{
Department of Physics,
The University of Tokyo,
Tokyo 113-0033,
Japan
}

\author{Keito Kasebayashi}
\affiliation{
Business Creation Sector R\&D Center,
MITSUI KINZOKU COMPANY,
LIMITED,
1333-2 Haraichi,
Ageo-shi,
Saitama,
362-0021,
Japan
}

\author{Hiroki Takahashi\orcidlink{0009-0006-5086-700X}}
\affiliation{
Business Creation Sector R\&D Center,
MITSUI KINZOKU COMPANY,
LIMITED,
1333-2 Haraichi,
Ageo-shi,
Saitama,
362-0021,
Japan
}

\author{Yu-ichiro Matsushita\orcidlink{0000-0002-9254-5918}}
\affiliation{
Quemix Inc.,
Taiyo Life Nihombashi Building,
2-11-2,
Nihombashi Chuo-ku, 
Tokyo 103-0027,
Japan
}

\affiliation{Quantum Materials and Applications Research Center,
National Institutes for Quantum Science and Technology (QST),
2-12-1 Ookayama, Meguro-ku, Tokyo 152-8550, Japan
}

\affiliation{
Department of Physics,
The University of Tokyo,
Tokyo 113-0033,
Japan
}

\date{\today}

\begin{abstract}
For treating correlated electronic systems on quantum computers,
we propose a quantum-classical hybrid scheme for dynamical mean-field theory (DMFT).
In the quantum part of the scheme,
we use modified quantum phase estimation (QPE) circuits suitable for the one-particle Green's function (GF) at a finite temperature so that we can extract spectral amplitudes and the excitation energies without knowing the excitation channel invoked at each measurement.
In the classical part of the scheme, we adopt an approach that estimates reasonably the GF based on the data collected from the QPE sampling.
We dub the approach the QPE averaged over variable grids (QAVG),
that may help one to reconstruct the GF via optimization of trial parameters and modeling the probability distributions for various settings of the QPE circuits.
We apply the QAVG-DMFT scheme to SrVO$_3$ to demonstrate its validity via numerical simulations.
\end{abstract}

\maketitle 

\section{Introduction}
\label{sec:introduction}

Electronic-structure calculations based on density functional theory (DFT) for classical computers have proven highly successful in providing quantitatively accurate predictions and explanations of the properties of a wide range of electronic systems.
In this framework, the complex behavior of interacting electrons is typically reformulated into an auxiliary problem of non-interacting electrons subject to an effective mean-field-like potential, which leads to the well-known Kohn–Sham (KS) equations.
Despite these successes, DFT-based methods often fail to capture even qualitative aspects of material properties in strongly correlated systems.
To overcome such limitations, numerous approaches have been developed and proposed in the literature.
One of those for reliable description of correlated systems are
the dynamical mean-field theory (DMFT) in conjunction with DFT,
denoted as DFT+DMFT approaches \cite{bib:3127}.
A typical one of them constructs a second-quantized isolated system from the KS orbitals near the Fermi level and solve a corresponding Anderson impurity model (AIM).

DMFT calculations using real quantum computers have been already reported.
They include DMFT for the Hubbard--Holstein model \cite{bib:6966} by employing the Krylov variational quantum algorithm (KVQA) method \cite{bib:7240,bib:7241},
DFT+DMFT for a copper oxide \cite{bib:6965} by employing the quantum equation of motion (qEOM) method \cite{bib:6994,bib:7242},
and DMFT for the Hubbard model \cite{bib:7117} by employing the cumulant expansion method \cite{bib:7243}.
While the approaches adopted in these earlier studies were chosen according to their applicability to current quantum hardware,
we adopt the quantum phase estimation (QPE) based on the quantum Fourier transform (QFT) \cite{bib:4825, bib:4826, Nielsen_and_Chuang} as the central technique in our scheme, as will be described.
It is because of the rapid growth in the number of high-fidelity qubits on hardware \cite{bib:6326, bib:7108, bib:7301} and the recent achievements for logical qubits \cite{bib:6948, bib:7441, bib:7111, bib:7246} toward the era of fault-tolerant quantum computation (FTQC).

Modern DMFT calculations as entirely classical computation often adopt versions of quantum Monte Carlo (QMC) methods as impurity solvers to get the one-particle Green's function (GF).
On the other hand, DMFT with full configuration interaction (FCI) solvers, or equivalently exact-diagonalization solvers, play an important role in the present study
since we establish the quantum-classical hybrid scheme based on the FCI-based classical scheme, that introduces a finite number of bath sites whose parameters are updated iteratively in the DMFT feedback loop.
The quantum part in our scheme begins with preparation of the Gibbs state as a many-qubit state, which then undergoes modified QPE circuits equipped with excitation partial circuits.
The sampled data let us estimate the finite-temperature GF without knowing the excitation channel invoked in each measurement and the energy eigenvalues.
For estimating the GF from those data,
we extend the QPE averaged over variable grids (QAVG) approach,
which was developed originally for a zero-temperature system \cite{QAVG_for_Fe2C5-CO} as post processing,
to finite-temperature cases.
We optimize trial parameters and model probability distributions so that the sampled data are reproduced accurately. 
The estimated GF is then fed into the DMFT loop to continue the iterations.
We apply the new DMFT scheme to SrVO$_3$ to demonstrate its validity via numerical simulations.

\section{Methods}

\subsection{One-particle GFs}

Before going to the DFT+DMFT scheme, we recapitulate the definition and useful expressions for the one-particle GF of a generic many-electron system.
Let $\mathcal{H}$ be the second-quantized Hamiltonian of the system composed of $n_{\mathrm{sorb}}$ spin orbitals.
In this subsection, we assume that the effect of chemical potential, or equivalently the Fermi level, is already taken into account in the one-body part of $\mathcal{H}.$

The one-particle GF in frequency domain is written as
\begin{align}
    G_{m m'} (z)
    &=
        G_{m m'}^{(e) } (z)
        +
        G_{m m'}^{(h) } (z)
\end{align}
for a complex frequency $z.$
$m$ and $m'$ are indices of spin orbitals.
$G^{(e) }$ and $G^{(h) }$ are the electron and hole parts, respectively, given by \cite{bib:3066}
\begin{align}
    G_{m m'}^{(\xi) } (z)
    &=
        \frac{1}{\mathcal{Z}}
        \sum_{\lambda_0}
            e^{-\beta E_{\lambda_0} }
            G_{m m'}^{(\xi) } (z | \lambda_0)
    \
    (\xi = e, h)
    ,
    \label{G_imag_freq_sum_partial_G}
\end{align}
where $\mathcal{Z} \equiv \mathrm{Tr} e^{-\beta \mathcal{H}}$ is the partition function for an inverse temperature $\beta.$
The summation in Eq.~(\ref{G_imag_freq_sum_partial_G}) is taken over each energy eigenstate $| \Psi_{\lambda_0} \rangle$ as an initial state of transition,
whose energy eigenvalue is $E_{\lambda_0}.$
It is reasonable, however, in a practical calculation to introduce a threshold $\varepsilon_{\mathrm{B}}$ for the Boltzmann factors $e^{-\beta E_{\lambda_0} }$ such that only the excitations whose Boltzmann factors are larger than $\varepsilon_{\mathrm{B}}$ are incorporated into the GF for avoiding too many excitation channels.
\begin{align}
    G_{m m'}^{(e) } (z | \lambda_0)
    &\equiv
        \sum_{\lambda}
        \frac{
            \langle \Psi_{\lambda_0} | a_m | \Psi_{\lambda} \rangle
            \langle \Psi_{\lambda} | a_{m'}^\dagger | \Psi_{\lambda_0} \rangle
        }{ z - (E_{\lambda} - E_{\lambda_0}) }
    \label{def_partial_G_e}
\end{align}
is the Lehmann representation \cite{stefanucci2013nonequilibrium} of the partial GF responsible for the electron excitations from $| \Psi_{\lambda_0} \rangle.$
$a_m^\dagger$ and $a_m$ are the creation and annihilation operators, respectively, for an electron at the $m$th spin orbital.
The summation in the equation above is taken over the energy eigenstates whose electron number is more by one than $| \Psi_{\lambda_0} \rangle.$
\begin{align}
    G_{m m'}^{(h) } (z | \lambda_0)
    \equiv
        \sum_{\lambda}
        \frac{
            \langle \Psi_{\lambda_0} | a_{m'}^\dagger | \Psi_{\lambda} \rangle
            \langle \Psi_{\lambda} | a_m | \Psi_{\lambda_0} \rangle
        }{ z - ( E_{\lambda_0} - E_{\lambda} )}
    \label{def_partial_G_h}
\end{align}
is the partial GF responsible for the hole excitations,
where the summation is taken over the energy eigenstates whose electron number is less by one than $| \Psi_{\lambda_0} \rangle.$
The partial GF for $\xi$ excitation clearly satisfies
$G^{(\xi)}_{m' m} (z) = G^{(\xi)}_{m m'} (z^*)^*.$

The density of states (DOS) of a many-electron system for the hole and electron parts are experimentally obtained via photoelectron spectroscopy (PES) and inverse PES processes, respectively \cite{bib:4070,bib:4165}.
They are to be compared under certain assumptions with the calculated DOS defined as \cite{bib:7463,bib:4473,bib:4483,bib:4516,bib:4575}
\begin{align}
    \rho (\omega)
    =
        -
        \frac{1}{\pi}
        \mathrm{Im tr}
        G (\omega + i \delta)
    ,
\end{align}
where $\omega$ is an excitation energy and $\delta$ is an infinitesimally small real constant.
For the $\xi$ excitation between an initial state $| \Psi_{\lambda_0} \rangle$ and a final state $| \Psi_{\lambda} \rangle,$
we define the transition matrix $B^{(\lambda_0 \to \lambda, \xi)}$ via
\begin{gather}
    B_{m m'}^{(\lambda_0 \to \lambda, e)}
    \equiv
        \frac{e^{-\beta E_{\lambda_0} } }{\mathcal{Z}}
        \langle \Psi_{\lambda_0} | a_m | \Psi_\lambda \rangle
        \langle \Psi_\lambda | a_{m'}^\dagger | \Psi_{\lambda_0} \rangle
    \label{def_transition_mat_for_G_e}
\end{gather}
and
\begin{gather}
    B_{m m'}^{(\lambda_0 \to \lambda, h)}
    \equiv
        \frac{e^{-\beta E_{\lambda_0} } }{\mathcal{Z}}
        \langle \Psi_{\lambda_0 } | a_{m'}^\dagger | \Psi_\lambda \rangle
        \langle \Psi_\lambda | a_m | \Psi_{\lambda_0 } \rangle
    .
    \label{def_transition_mat_for_G_h}
\end{gather}
These matrices are clearly Hermitian:
$
B_{m' m}^{(\lambda_0 \to \lambda, \xi)}
=
B_{m m'}^{(\lambda_0 \to \lambda, e) *}
.
$

Since we will work with QPE sampling in the present study,
we define the spectral matrix $S^{(\xi)}$ via
\begin{align}
    S^{(\xi)}_{m m'} (\varepsilon)
    &\equiv
        \sum_{\lambda_0, \lambda}
            B_{m m'}^{(\lambda_0 \to \lambda, \xi)}
            \delta_{E_{\lambda} - E_{\lambda_0}, \varepsilon}
    ,
    \label{avr_spectra_using_QPE:def_spectral_mat}
\end{align}
where the Kronecker delta is for the excitation energy
from $| \Psi_{\lambda_0} \rangle$ to $| \Psi_{\lambda} \rangle.$
The argument $\varepsilon$ is a possible discrete value represented by the ancillae used in the QPE circuit.
For a case where a large number of ancillae are used in the QPE circuit,
we can write the electron GF approximately as
\begin{align}
    G_{m m'}^{(e)} (z)
    &\approx
        \sum_{\varepsilon}
        \frac{S^{(e)}_{m m'} (\varepsilon)}{ z - \varepsilon }
    \label{avr_spectra_using_QPE:GF_el_using_kronecker_delta}
\end{align}
and the hole GF as
\begin{align}
    G_{m m'}^{(h)} (z)
    &\approx
        \sum_{\varepsilon}
        \frac{S^{(h)}_{m m'} (\varepsilon)}{ z + \varepsilon }
    .
    \label{avr_spectra_using_QPE:GF_hole_using_kronecker_delta}
\end{align}

\subsection{DFT+DMFT using a quantum computer}

\begin{figure}
\begin{center}
\includegraphics[width=7cm]{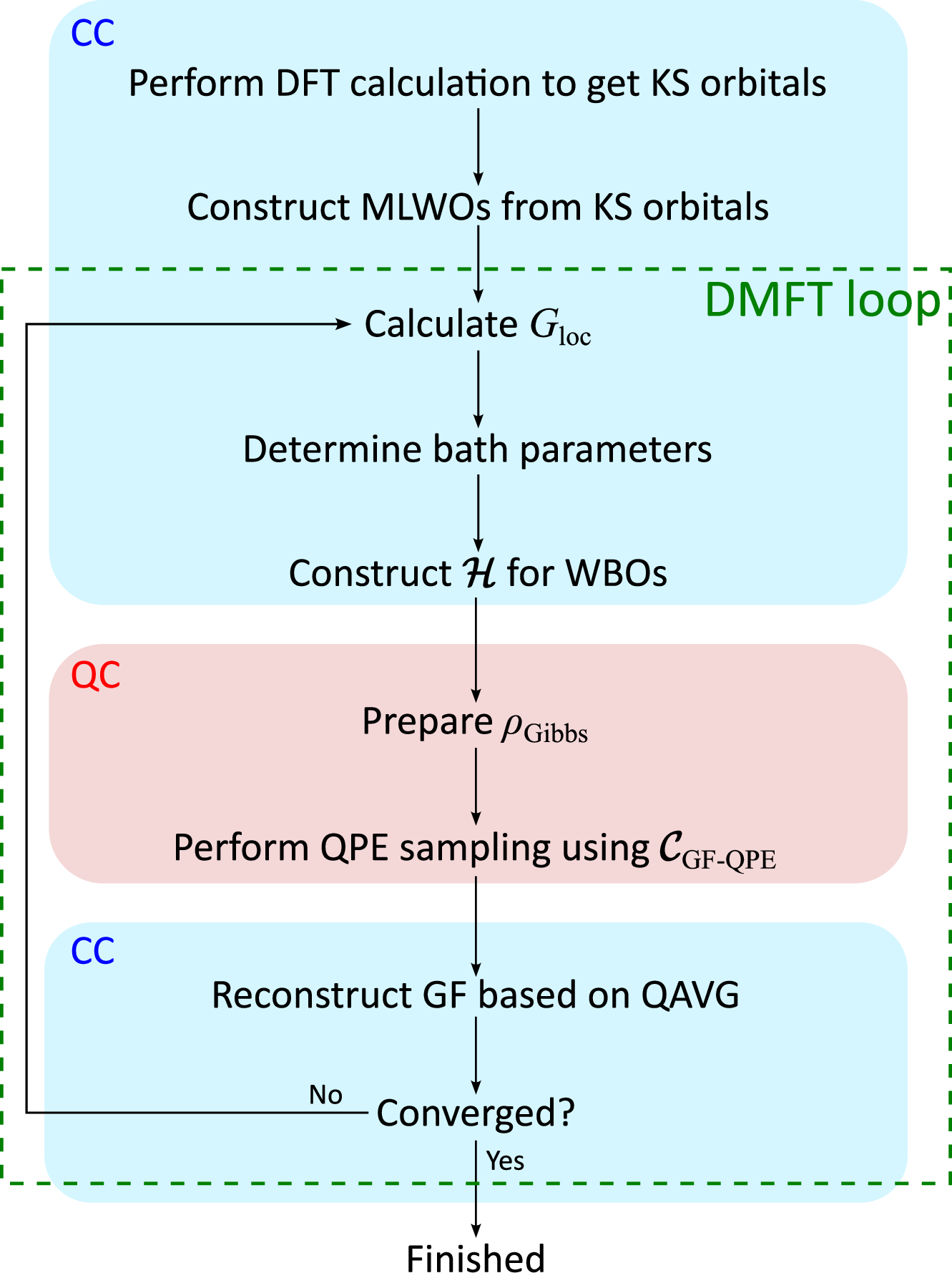}
\end{center}
\caption{
Flowchart of DFT+DMFT calculation based on QAVG in the present study.
Items in the blue regions are carried out on a classical computer (CC),
while those in the red region are on a quantum computer (QC).
This scheme is the same as an ordinary FCI-based DFT+DMFT scheme except for the QC part.
}
\label{fig:dft_dmft_qavg_flowchart}
\end{figure}

Here we describe our scheme for a DFT+DMFT calculation based on QAVG.
The flowchart of the scheme is provided in
Fig.~\ref{fig:dft_dmft_qavg_flowchart}.

We assume that a self-consistent DFT calculation for a target periodic system and subsequent construction of the MLWOs \cite{bib:4596} on a classical computer has been done.
Let $n_{\mathrm{wan}}$ be the number of constructed MLWOs,
which constitute the band manifold around the Fermi level.
Amongst them, $n_{\mathrm{corr}}$ MLWOs are chosen to be involved in the DFMT calculation.
The lattice GF for $n_{\mathrm{wan}}$ MLWOs is defined as
\begin{align}
    G_{\mathrm{latt} } (\boldsymbol{k}, z)
    \equiv
        \frac{1}{
        z + \mu
        -
        H_{\mathrm{KS}} (\boldsymbol{k})
        -
        \Sigma_{\mathrm{loc}} (z)
        }
    \label{DFT_plus_DMFT:def_GF_latt_k}
\end{align}
for a wave vector $\boldsymbol{k}$ and a complex frequency $z.$
$H_{\mathrm{KS}} (\boldsymbol{k})$ is constructed from the KS Hamiltonian matrix in the MLWO representation and $\Sigma_{\mathrm{loc}}$ is the self-energy, which is assumed to be local, that is, independent of the wave vector.
$\mu$ is the chemical potential which should be tuned iteratively during the DMFT loop so that the desired occupancy of the correlated bands in the periodic system is achieved.
We set $\Sigma_{\mathrm{loc}}$ to zero in the first iteration of the loop,
while it can be nonzero in the subsequent iterations.
The reciprocal-space GF in Eq.~(\ref{DFT_plus_DMFT:def_GF_latt_k}) and the real-space GF for lattice vectors $\boldsymbol{R}$ are related via Fourier transform.
In particular,
\begin{align}
    G_{\mathrm{latt} } (\boldsymbol{R} = \boldsymbol{0}, z)
    =
        \frac{1}{N_{\boldsymbol{k}}}
        \sum_{\boldsymbol{k}}
        G_{\mathrm{latt} } (\boldsymbol{k}, z)
    \equiv
    G_{\mathrm{loc} } (z)
    \label{DFT_plus_DMFT:def_GF_latt_R0}
\end{align}
is called the local GF, responsible for the interactions within a unit cell.
$N_{\boldsymbol{k}}$ is the number of sampled $k$ points.

The DMFT scheme treats the isolated correlated system consisting of the $n_{\mathrm{corr}}$ correlated MLWOs and $n_{\mathrm{bath}}$ bath orbitals.
The dynamical mean field $\mathcal{G}_0,$ or equivalently the Weiss field,
for the correlated orbitals is defined from the local GF and the self-energy as
\begin{align}
    \mathcal{G}_0 (z)^{-1}
    \equiv
        P_{\mathrm{corr}}
        \left(
            G_{\mathrm{loc} } (z)^{-1}
            +
            \Sigma_{\mathrm{loc}} (z)
        \right)
        P_{\mathrm{corr}}
    ,
\end{align}
where $P_{\mathrm{corr}}$ is the projection matrix onto the subspace spanned by the $n_{\mathrm{corr}}$ MLWOs.

The parameters that characterize $n_{\mathrm{bath}}$ orbitals are determined by solving a nonlinear optimization problem so that the difference between the hybridization matrix $\Delta (i \omega_n)$ calculated from $\mathcal{G}_0$ and that expressed by bath parameters is minimized.
$\omega_n \equiv (2 n + 1) \pi/\beta \ (n = 0, \pm 1, \pm 2, \dots)$ are the fermionic Matsubara frequencies for an inverse temperature $\beta.$
We set $\beta^{-1}$ to the temperature at which the original periodic system is.
We refer to the correlated MLWOs together with bath orbitals as the Wannier and bath orbitals (WBOs) in what follows.
In addition to the transfer integrals and the interaction parameters 
$U, U_0,$ and $J$ \cite{bib:7218, bib:7027} among the correlated MLWOs,
the optimal bath parameters let us to construct the second-quantized Hamiltonian $\mathcal{H}$ of the isolated system consisting of $n_{\mathrm{sorb}} \equiv 2 (n_{\mathrm{corr}} + n_{\mathrm{bath}})$ WBOs, where $2$ comes from the spin degrees of freedom.

Having constructed the Hamiltonian for the AIM,
we then use a quantum computer to collect data for obtaining the GF of the isolated system instead of ordinary FCI-based DMFT schemes on classical computers.
Specifically, we prepare the Gibbs state $\rho_{\mathrm{Gibbs}}$ as a many-qubit state,
which we input to modified QPE circuits $\mathcal{C}_{\mathrm{GF-QPE}}$ responsible for the individual components of GF and perform repeated measurements.
The outcomes as histograms are used for a QAVG calculation on a classical computer.
The generic scheme of QAVG will be described in Sect.~\ref{sec:qpe_and_qavg}.
The explicit construction of the modified QPE circuits will be described in Sect.~\ref{sec:circuits_for_GF_at_nonzero_temp}.
The GF from the QAVG calculation allows us to get the self-energy $\Sigma_{\mathrm{corr}}$ of the correlated MLWOs from the Dyson equation.
We identify it as the self-energy $\Sigma_{\mathrm{loc}}$ in the periodic system,
which we substitute into Eq.~(\ref{DFT_plus_DMFT:def_GF_latt_k}) to update the lattice GF with tuning the chemical potential $\mu.$

\subsection{QPE and generic QAVG}
\label{sec:qpe_and_qavg}

\subsubsection{QFT-based QPE}

Among various versions of QPE,
we denote the standard QFT-based QPE \cite{bib:4825, bib:4826, Nielsen_and_Chuang} simply as the QPE in the present study.

The protocol of QPE for a Hermitian operator $\mathcal{A}$
is designed so that,
when any one $| a \rangle$ of the eigenstates of $\mathcal{A}$ is input to the QPE circuit,
repeated measurements on ancillae provide an estimation of the eigenvalue $a.$
The accuracy of estimation depends not only on the number $n_{q \mathrm{val}}$ of the ancillae for representing the eigenvalue in binary representation,
but also on the grid spacing $1/t_0$ and the origin $a_{\mathrm{orig}}$ for representing $N_{\mathrm{val}} \equiv 2^{n_{q \mathrm{val}}}$ possible estimators.
In the case where $a - a_{\mathrm{orig}}$ is not strictly equal to any one of the values that can be expressed by bit strings of length $n_{q \mathrm{val}},$
which is quite likely in practical situations,
the histogram of measurement outcomes have multiple high frequencies at values near $a - a_{\mathrm{orig}}$ modulo $N_{\mathrm{val}}/t_0.$
This phenomenon is known as the spectral leakage.
As for the implementation, the unitary
$
U
\equiv
\exp
( i 2 \pi (\mathcal{A} - a_{\mathrm{orig}}) t_0/ N_{\mathrm{val}} )
$
and its powers are controlled by the ancillae and act on the input state.
For details, see a textbook \cite{Nielsen_and_Chuang}.

\subsubsection{Generic QAVG}

\begin{figure*}
\begin{center}
\includegraphics[width=13cm]{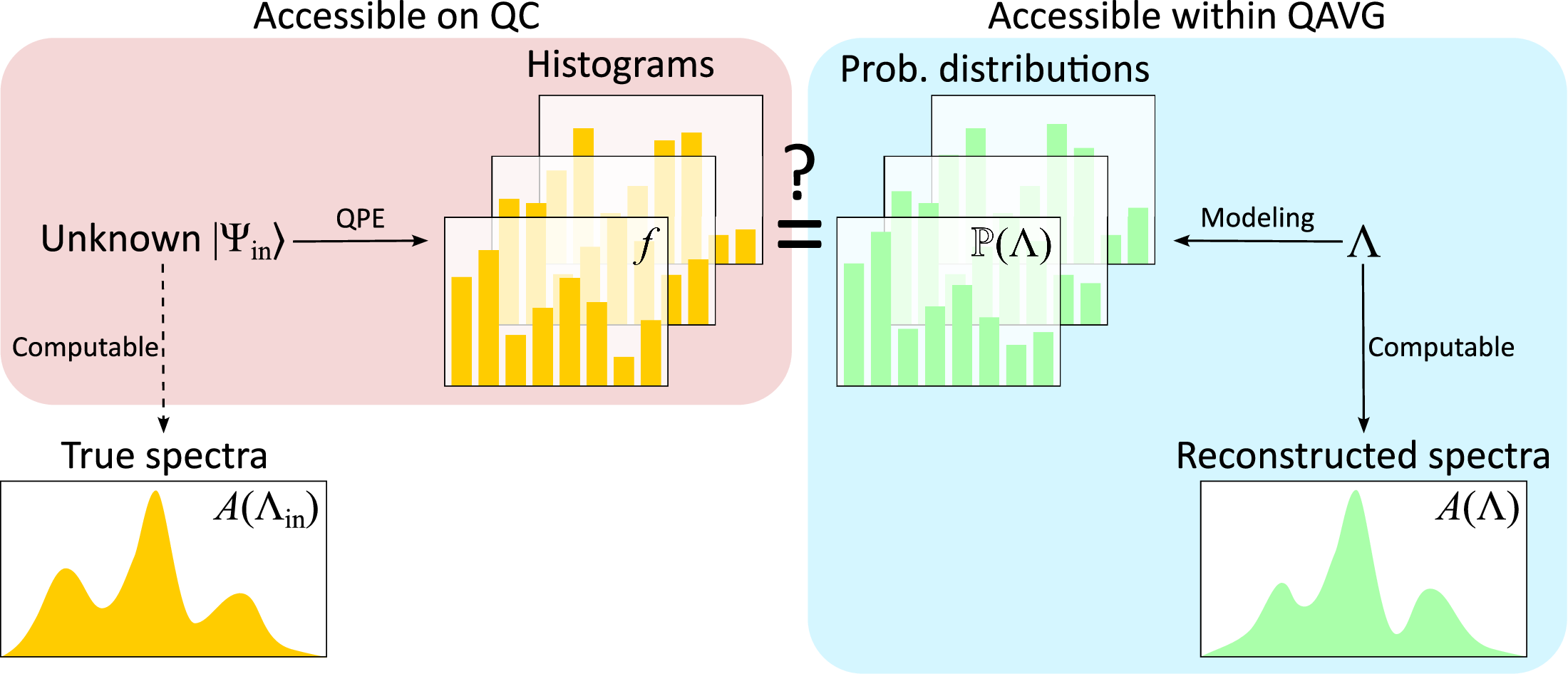}
\end{center}
\caption{
Illustration of QAVG.
Although we can prepare the input state $| \Psi_{\mathrm{in}} \rangle$ on a quantum computer,
its expression and the values of its related quantities $\Lambda_{\mathrm{in}}$ are unknown.
The true spectra $A (\Lambda_{\mathrm{in}})$ is computable in principle since its expression is known.
It is, however, actually impossible to get it due to the unknown $\Lambda_{\mathrm{in}}.$
On the other hand, what we can obtain directly are QPE histograms $f$ by performing a finite number of measurements for various QPE settings.
After collecting the histograms as data,
the QAVG scheme starts from defining trial parameters $\Lambda$ and model the probability distributions $\mathbb{P} (\Lambda)$ so that they describe QPE measurements reasonably.
The scheme then finds the optimal trial parameters $\Lambda_{\mathrm{opt}}$ to minimize the discrepancy between the data and the modeled probability distributions.
We expect that the reconstructed spectra $A (\Lambda_{\mathrm{opt}})$ predicts well what we really want, $A (\Lambda_{\mathrm{in}}).$ 
}
\label{fig:qavg_illustration}
\end{figure*}

The generic scheme of QAVG is illustrated in Fig.~\ref{fig:qavg_illustration}.
Let us consider an unknown many-qubit state $| \Psi_{\mathrm{in}} \rangle$ for which we want to calculate a related quantity $A$ such as the spectral function or the GF.
The phrase `unknown' means that we can prepare the state by obeying some definite protocol,
but we do not know an expression that describes the state succinctly.
QAVG provides a way for inferring $A$ that is essentially unrelated to QPE based on the results of QPE sampling. 
To avoid the biases originating from a specific combination
$(t_0, a_{\mathrm{orig}})$ of the scaling parameter and the origin,
we propose to use QPE circuits specified by $n_{\mathrm{setting}}$ settings
$\{ (t_0^{(p)}, a_{\mathrm{orig}}^{(p)} ) \}_{p = 0}^{n_{\mathrm{setting}} - 1 }$
of the circuit parameters.

We construct the QPE circuit $\mathcal{C}_p$ specified by each of the settings and input $| \Psi_{\mathrm{in}} \rangle$ to it.
Let us denote the quantities associated with $| \Psi_{\mathrm{in}} \rangle,$
such as the excitation energies and the transition amplitudes, etc., 
collectively by $\Lambda_{\mathrm{in}}.$
The measurement outcomes from the $n_{q \mathrm{val}}$ ancillae comprising $\mathcal{C}_p$ obey some probability distribution $\mathbb{P}^{(p)} (\Lambda_{\mathrm{in}}),$
where
$\mathbb{P}_j^{(p)} (\Lambda_{\mathrm{in}})$ is the probability for observing an integer $j$ as a bit string of length $n_{q \mathrm{val}}.$
This distribution of course satisfies the sum rule
$
\sum_{j = 0}^{N_{\mathrm{val}} - 1}
\mathbb{P}^{(p)}_j (\Lambda_{\mathrm{in}})
=
1
.
$

Here we introduce a crucial assumption for formulating QAVG:
the expressions for $A (\Lambda_{\mathrm{in}})$ and $\mathbb{P}^{(p)} (\Lambda_{\mathrm{in}})$ as functions of $\Lambda_{\mathrm{in}}$ are known despite the values of $\Lambda_{\mathrm{in}}$ being unknown. 
What we obtain in practice is a histogram $f^{(p)}$ from a finite number of measurements, which may contain statistical noises.
After finishing all the measurements for all the settings,
we have the data $\{ f^{(p)}_j \}_{p, j}.$
QAVG aims to determine $\Lambda_{\mathrm{in}}$ by regarding them as independent trial parameters $\Lambda$ so that the computed probability distributions reproduce the obtained histograms as accurately as possible.
To this end,
by averaging the discrepancies $D$ between the probability distributions derived from $\Lambda$ and the data over the variable QPE grids,
we define the cost function
\begin{align}
    F
    (\Lambda)
    &\equiv
        \frac{1}{n_{\mathrm{setting}}}
        \sum_{p = 0}^{n_{\mathrm{setting}} - 1}
        D
        \left(
            \mathbb{P}^{(p)}
            (\Lambda)
            ,
            f^{(p)}
        \right)
        .
\end{align}
Any functional form of $D$ is admissible as long as it quantifies the closeness between two input distributions in some sense,
such as the $L_1$ distance \cite{Nielsen_and_Chuang}.
We define in the present study the nonuniform-weight $L_1$ distance as
\begin{align}
    D (h, h')
    \equiv
        \sum_{j = 0}^{N_{\mathrm{val}} - 1}
        g_j \frac{| h_j - h'_j |}{2}
    \label{fig:def_nonuniform_weight_L1}
\end{align}
for two distributions $h$ and $h'.$
$g_j$ is the weight of $j$th grid point satisfying $\sum_j g_j = 1.$
The value of $D$ falls between $0$ and $1.$
QAVG adopts the trial parameters that minimize the cost function as the optimal ones:
\begin{align}
    \Lambda_{\mathrm{opt}}
    =
        \argmin_{\Lambda} F (\Lambda) 
        .
\end{align}
We expect that the optimal parameters not only reproduce the data,
but also provide $A (\Lambda_{\mathrm{opt}})$ as a reliable prediction of $A (\Lambda_{\mathrm{in}}).$
Although the descriptions provided just above assumed the input state to be pure,
QAVG is also applicable to a mixed state.

When an adopted form of $D$ satisfies the triangle inequality,
we can find an upper bound of the discrepancy between the probability distribution $\mathbb{P}_{\mathrm{noiseless}}^{(p)}$ that would be attained if the actual circuit $\mathcal{C}_p$ was noiseless and the computed probability distribution $\mathbb{P}^{(p)} (\Lambda)$ for given trial parameters.
Specifically, we can write
 \begin{gather}
    D
    \left(
        \mathbb{P}^{(p)} (\Lambda),
        \mathbb{P}_{\mathrm{noiseless}}^{(p)}
    \right)
    \nonumber \\
    \leq
    D
    \left(
        \mathbb{P}^{(p)} (\Lambda),
        \mathbb{P}^{(p)} (\Lambda_{\mathrm{opt}})
    \right)
    +
    D
    \left(
        \mathbb{P}^{(p)} (\Lambda_{\mathrm{opt}}),
        f^{(p)}
    \right)
    \nonumber \\
    +
    D
    \left(
        f^{(p)},
        \mathbb{P}_{\mathrm{noisy}}^{(p)}
    \right)
    +
    D
    \left(
        \mathbb{P}_{\mathrm{noisy}}^{(p)},
        \mathbb{P}_{\mathrm{noiseless}}^{(p)}
    \right)
    .
    \label{DFT_plus_DMFT:upper_bound_of_discrepancy}
\end{gather}
We can interpret the meaning of each term on the right-hand side of Eq.~(\ref{DFT_plus_DMFT:upper_bound_of_discrepancy}) as follows.
The first term represents the optimization error for the trial parameters.
The second term represents the parametrization error, that is,
if it is nonzero, the observed histogram $f$ is not reproduced exactly even by the fully optimized trial parameters.
The third term represents the statistical error due to the finite number of measurements on the noisy circuit.
The circuit $\mathcal{C}_p$ leads to the probability distribution $\mathbb{P}_{\mathrm{noisy}}^{(p)}$ for an infinite number of measurements.
The fourth term represents all kinds of errors other than the statistical error.
The central idea of QAVG is to make the second term as small as possible.
How small the first term can be made depends on the choice of numerical algorithms for minimization.
The contribution from the fourth term, on the other hand, is decreased mainly by hardware-level improvement.

\subsection{QPE Circuits for GF at a finite temperature}
\label{sec:circuits_for_GF_at_nonzero_temp}

\subsubsection{Circuit construction}

There already exist approaches for obtaining the one-particle GF \cite{bib:5005} and the linear-response functions \cite{bib:5163} of a second-quantized many-electron system at zero temperature via QPE sampling.
Those approaches assume that the energy eigenvalue of ground states is known for constructing the histograms from the measurements.
They are, however, not available in the DMFT scheme for the present study since, as described above, the GF at the discrete Matsubara frequencies whose spacing is proportional to the temperature are assumed.
Furthermore, transitions from excited states contribute to the GF and one is thus not able to identify the initial state of a transition observed at each QPE measurement.

To circumvent these difficulties,
we propose a circuit applicable to a finite-temperature system.
The building block of the circuit is the RTE operator
\begin{align}
    U_{\mathrm{RTE}} (s)
    \equiv
        \exp
        \left(
            -i 2 \pi \mathcal{H}
            \frac{t_0 s}{N_{\mathrm{val}}}
        \right)
    \label{avr_spectra_using_QPE:def_unitary_in_QPE}
\end{align}
as well as in the ordinary QPE.
We assume that practical circuit implementation of this unitary is possible by employing various techniques such as the Suzuki--Trotter decomposition for an arbitrary $s.$ 
To obtain the one-particle GF at a finite temperature by performing QPE sampling,
we arrange the controlled RTE gates [see Fig.~\ref{fig:circuit_gf_for_gibbs}(a)] to build the circuit $\mathcal{C}_{\mathrm{GF-QPE}},$ as shown in
Fig.~\ref{fig:circuit_gf_for_gibbs}(b).
$\mathcal{C}_{\mathrm{GF-QPE}}$ receives the Gibbs state 
$\rho_{\mathrm{Gibbs}}$ as an input to the data register.
There are various quantum algorithms for performing genuine or effective Gibbs sampling proposed so far \cite{bib:7221,bib:4959,bib:7223,bib:7220,bib:5737,bib:7222,bib:7224,bib:7205,bib:7214}.
We do not assume any specific algorithm as to how $\rho_{\mathrm{Gibbs}}$ is prepared.

\subsubsection{Excitation}

The explicit construction of the excitation circuit 
$\mathcal{C}_{\mathrm{exc}}$ contained in $\mathcal{C}_{\mathrm{GF-QPE}}$
depends on the pair of WBOs representing the component of the GF.
For collecting data to evaluate the cost function in terms of the diagonal component $G_{m m}$ for the $m$th WBO,
we use the diagonal excitation circuit $\mathcal{C}_m$ as $\mathcal{C}_{\mathrm{exc}}.$
For the pair of $m$th and $m'$th WBOs $(m \ne m'),$ on the other hand,
we use the off-diagonal excitation circuit $\mathcal{C}_{m m'}.$ 
For details, see Appendix \ref{sec:excitation_circuits}.
In addition,
the measurement outcomes from the excitation circuits provide estimation $\gamma_{\mathrm{meas}, m m'}$ of the one-electron matrix element
$\gamma_{m m'} \equiv \langle a_m^\dagger a_{m'} \rangle,$
as described in Appendix \ref{sec:excitation_on_gibbs_state}.
The total number of shots for determining all the components of $\gamma$ with a specified accuracy $\epsilon$ and a failure probability $p_{\mathrm{fail}}$ needs to satisfy
$M_{\mathrm{tot}} \geq \mathcal{O} (n_{\mathrm{sorb}}^2/(\epsilon^2 p_{\mathrm{fail}}) ).$

\subsubsection{Channel-agnostic extraction of excitation energies}

When one of the energy eigenstates $| \Psi_{\lambda_0} \rangle$ together with the ancillary state $| + \rangle^{\otimes n_{q \mathrm{val}} }$ is input to $\mathcal{C}_{\mathrm{GF-QPE}},$
the controlled RTE gates before $\mathcal{C}_{\mathrm{exc}}$ give rise to the phase factor indicating $-E_{\lambda_0}.$
$\mathcal{C}_{\mathrm{exc}}$ then prepares a superposition $| \Psi_{\mathrm{exc}} \rangle$ of the energy eigenstates as explained above,
after which the controlled RTE gates give rise to the additional phase factor indicating $E_{\lambda}$ for each of the constituent energy eigenstates $| \Psi_{\lambda} \rangle.$
In short, $\mathcal{C}_{\mathrm{GF-QPE}}$ transforms the entire state as follows:
\begin{align}
    | \Psi_{\lambda_0} \rangle
    | + \rangle^{\otimes n_{q \mathrm{val}} }
    \rightarrow
    \sum_{\lambda}
        \langle \Psi_{\lambda} | \Psi_{\mathrm{exc}} \rangle
        | \Psi_{\lambda} \rangle
        | E_{\lambda} - E_{\lambda_0} \rangle_{n_{q \mathrm{val}}}
    ,
    \label{avr_spectra_using_QPE:circ_GF_QPE_for_energy_eigenstate}
\end{align}
where
$| E_{\lambda} - E_{\lambda_0} \rangle_{n_{q \mathrm{val}}}$
is the ancillary state whose major amplitudes of computational bases represent the binary representation of the excitation energy.
The origin $E_{\mathrm{orig}}$ of extracted phase together with the grid spacing $1/t_0$ can be set by tuning the gate parameters in the RTE unitaries.
Since the Gibbs state is a classical mixture of the energy eigenstates,
we can also find the final state when $\rho_{\mathrm{Gibbs}}$ undergoes $\mathcal{C}_{\mathrm{GF-QPE}}.$
For details, see Appendix \ref{sec:exc_energy_meas_for_Gibbs}.
We can thus obtain the histograms that approximate the GF in the limited resolution by accumulating the measurement outcomes.
The striking feature of this approach is that the histograms can be obtained without knowing the energy eigenvalues and the excitation channel invoked at each measurement.
This feature is in contrast to the previous QPE-based approach \cite{bib:5005}, which is applicable only to a zero-temperature case where the energy of ground state is known.

It is noted that various approaches \cite{bib:7256, bib:5570, bib:7257, bib:7236, bib:7258} have been proposed so far for estimating energy differences based on QPE.
Among them, the Bayesian approaches \cite{bib:5570, bib:7257, bib:7236} make use of binary measurement outcomes to obtain the difference in energy eigenvalues between two input pure states.
One of the nice features in those approaches compared to $\mathcal{C}_{\mathrm{GF-QPE}}$ is that they do not require implementation of controlled RTE gates. 
There also exists an approach for obtaining the finite-temperature GF based on binary measurements in imaginary-time domain \cite{bib:7112},
which is then classically Fourier transformed to the Matsubara GF at imaginary frequencies.
The GF for real frequencies can be obtained via analytic continuation.

\begin{figure*}
\begin{center}
\includegraphics[width=11cm]{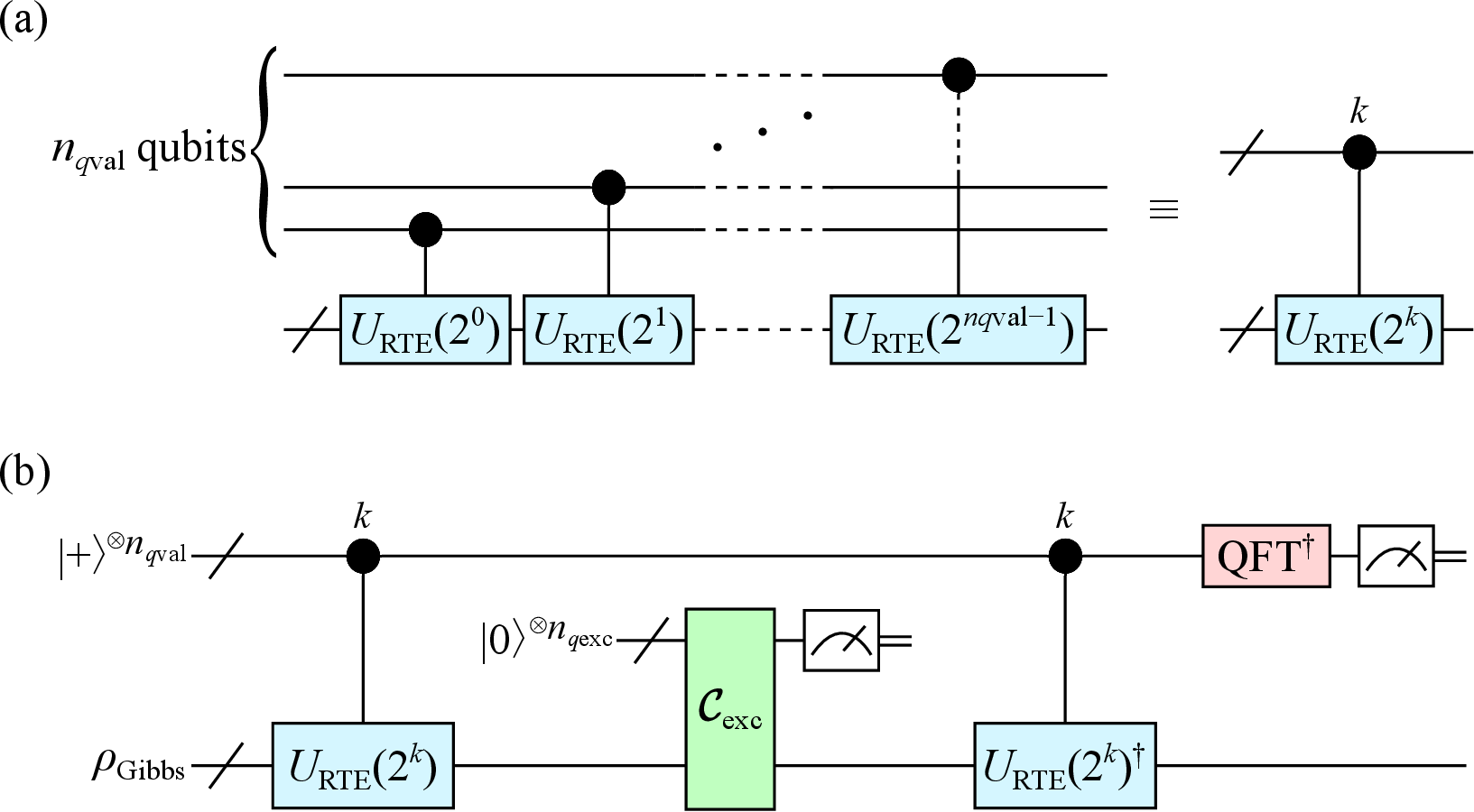}
\end{center}
\caption{
(a)
We denote the left circuit,
where $n_{q \mathrm{val}}$ RTE unitaries controlled by the ancillae with distinct amounts of time steps,
by the right circuit with a dummy variable $k$ for simplicity.
(b)
Circuit $\mathcal{C}_{\mathrm{GF-QPE}}$ for performing channel-agnostic QPE sampling to obtain the GF at a finite temperature.
$\mathcal{C}_{\mathrm{exc}}$ contained in $\mathcal{C}_{\mathrm{GF-QPE}}$ is an excitation circuit that causes an electron- or hole-excitation depending on each measurement outcome for $n_{q \mathrm{exc}}$ ancillae. 
}
\label{fig:circuit_gf_for_gibbs}
\end{figure*}

\subsection{QAVG for reconstructing GF}

\subsubsection{Cost functions}

Let $\{ f^{(p, m)}_{\xi j} \}_{\xi, j}$ be the histogram obtained from the repeated measurements on the ancillae for the QPE in $\mathcal{C}_{\mathrm{GF-QPE}}$ consisting of the diagonal excitation circuit $\mathcal{C}_m$ and the $p$th setting.
Also, let $\{ f^{(p, m m')}_{\xi \sigma j} \}_{\xi, \sigma, j}$ be that for $\mathcal{C}_{\mathrm{GF-QPE}}$ with the off-diagonal excitation circuit $\mathcal{C}_{m m'}.$ 
When we finish the measurements by using all the $\mathcal{C}_{\mathrm{GF-QPE}}$ circuits,
we will have the data $\{ f^{(p, m)}_{\xi j}, f^{(p, m m')}_{\xi \sigma j} \}_{p, m, m', \xi, \sigma, j}.$
We assume the histograms to be normalized as
\begin{align}
    \sum_{\xi = e,h}
    \sum_{j = 0}^{N_{\mathrm{val}} - 1}
        f^{(p, m)}_{\xi j}
    =
        1
\end{align}
and
\begin{align}
    \sum_{\xi = e,h}
    \sum_{\sigma = +,-}
    \sum_{j = 0}^{N_{\mathrm{val}} - 1}
        f^{(p, m m')}_{\xi \sigma j}
    =
        1
\end{align}
for each of $m, m',$ and $p.$

Although we have not decided what are trial parameters $\Lambda$ for QAVG at this point,
we define the cost function formally here. 
For the histogram obtained by using the $\mathcal{C}_m$ circuit, we define the partial cost functions via
\begin{align}
    F^{(m)}_{\xi} (\Lambda_\xi)
    &\equiv
        \frac{1}{n_{\mathrm{setting}} }
        \sum_{p = 0}^{n_{\mathrm{setting}} - 1}
        D
        \left(
            \mathbb{P}^{(p, m)}_{\xi}
            (\Lambda_\xi)
            ,
            f^{(p, m)}_{\xi}
        \right)
\end{align}
for each of $\xi = e, h.$
$\Lambda_\xi$ are the trial parameters responsible for the $\xi$ excitations.
$\Lambda_e$ and $\Lambda_h$ do not overlap and $\Lambda$ is the union of them. 
$\mathbb{P}^{(p, m)}_{\xi}$ is a probability distribution modeled by trial parameters.
We also define the partial cost functions for the $\mathcal{C}_{m m'}$ circuit via
\begin{gather}
    F^{(m m')}_{\xi} (\Lambda_\xi)
    \nonumber \\
    \equiv
        \frac{1}{2 n_{\mathrm{setting}} }
        \sum_{p = 0}^{n_{\mathrm{setting}} - 1}
        \sum_{\sigma = +, -}
        D
        \left(
            \mathbb{P}^{(p, m m')}_{\xi \sigma}
            (\Lambda_\xi)
            ,
            f^{(p, m m')}_{\xi \sigma}
        \right)
        .
\end{gather}
It is clear that $F^{(m)}_{\xi} (\Lambda_\xi) \leq 1$
and $F^{(m m')}_{\xi} (\Lambda_\xi) \leq 1.$
Since we are interested mainly in low-energy excitations,
we introduce the exponential weights of grid points for the nonuniform-weight $L_1$ distance in Eq.~(\ref{fig:def_nonuniform_weight_L1}) as
\begin{align}
    g_j
    \equiv
        \frac{
            \exp (- \tau_{\mathrm{dec}} j/t_0 )
        }{
            \sum_{j'}
            \exp (- \tau_{\mathrm{dec}} j'/t_0 )        
        }
    ,
\end{align}
where $\tau_{\mathrm{dec}}$ is the decay rate.
We assign higher weights to lower-energy excitations since we would like to reconstruct the GF accurately particularly in low-energy regime.

By collecting the contributions, we define the total cost function for $\xi$ excitation via
\begin{gather}
    F_{\xi} (\Lambda_{\xi})
    \equiv
        \sum_m
        w_{\xi}^{(m)}        
        F^{(m)}_{\xi}
        (\Lambda_{\xi})
        +
        \sum_{m \ne m'}
        w_{\xi}^{(m m')}
        F^{(m m')}_{\xi}
        (\Lambda_{\xi})
        .
    \label{avr_spectra_using_QPE:def_cost_func}
\end{gather}
$w_{\xi}^{(m)}$ and $w_{\xi}^{(m m')}$ are the weights assigned to the $\mathcal{C}_m$ and $\mathcal{C}_{m m'}$ circuits, respectively,
which we require satisfy
$
\sum_m
w_{\xi}^{(m)}        
+
\sum_{m \ne m'}
w_{\xi}^{(m m')}
=
1
.
$
We have thus $F_{\xi} (\Lambda_{\xi}) \leq 1.$
The simplest definitions of the weights are
$w_{\xi}^{(m)} \equiv 1/n_{\mathrm{sorb}}^2$ and
$w_{\xi}^{(m m')} \equiv 1/n_{\mathrm{sorb}}^2.$
Alternatively,
based on the idea that
more frequent excitations should provide larger contributions to an inferred GF,
we define the following weights according to the number of observed $\xi$ excitations:
\begin{align}
    w_{\xi}^{(m)}
    \equiv
        \frac{1}{n_{\mathrm{sorb}} }
        \frac{M_{\xi}^{(m)} }{\sum_{m'} M_{\xi}^{(m')} }
    \label{avr_spectra_using_QPE:def_circ_weight_diag}
\end{align}
and
\begin{align}
    w_{\xi}^{(m m')}
    \equiv
        \left(
            1
            -
            \frac{1}{n_{\mathrm{sorb}} }
        \right)
        \frac{
            \sum_{\sigma} M_{\xi \sigma}^{(m m')}
        }{
            \sum_{m'' \ne m''', \sigma'} M_{\xi \sigma' }^{(m'' m''')}
        }
        .
    \label{avr_spectra_using_QPE:def_circ_weight_off_diag}
\end{align}
$M_\xi^{(m)}$ is the number of $\xi$ excitations observed on $\mathcal{C}_m$
and
$M_{\xi \sigma}^{(m m')}$ is that for $\xi \sigma$ excitations observed on $\mathcal{C}_{m m'}.$
We adopt the weights in
Eqs.~(\ref{avr_spectra_using_QPE:def_circ_weight_diag}) and
(\ref{avr_spectra_using_QPE:def_circ_weight_off_diag}) for the present study.
We can relate these definitions to the measured one-electron matrix elements $\gamma_{\mathrm{meas}},$
as described in Appendix~\ref{sec:Circuit_weights_for_gamma_meas}.

\subsubsection{Trial parameters based on natural orbitals}

Since the target systems for DMFT calculations in the present study are isolated and non-relativistic,
the transition amplitudes can be assumed to be real without loss of generality.
As described in Appendix \ref{sec:natural_orb_repr},
the GF in natural-orbital (NO) representation can be expressed by the transition amplitudes that form orthonormalized systems. 
We therefore work with the NO representation rather than the WBO representation for modeling the GF via trial parameters.

Given the expressions of GF in the NO representation in
Eqs.~(\ref{avr_spectra_using_QPE:gf_in_nat_orb_repr_e}) and
(\ref{avr_spectra_using_QPE:gf_in_nat_orb_repr_h}),
we propose the reconstructed GF $\widetilde{G}_{\mathrm{rec}} (z; \Lambda)$ with
\begin{gather}
    \widetilde{G}_{\mathrm{rec}, \nu \nu'}^{(e)}
    (z; \Lambda_e)
    \equiv
        \sqrt{1 - n_{\mathrm{meas}, \nu}}
        \sqrt{1 - n_{\mathrm{meas}, \nu'}}
    \cdot
    \nonumber \\
    \cdot
        \sum_{\ell = 0}^{n_{\mathrm{ch}} - 1 }
        \int_{-\infty}^\infty
        dE
        \rho_{e \ell} (E - \varepsilon_{e \ell})
        \frac{
            v^{(\nu)}_{e \ell}
            v^{(\nu')}_{e \ell}
        }{z - E }
    \label{avr_spectra_using_QPE:gf_rec_in_NO_using_DOS_e}
\end{gather}
for the electron part and 
\begin{gather}
    \widetilde{G}_{\mathrm{rec}, \nu \nu'}^{(h)}
    (z; \Lambda_h)
    \equiv       
        \sqrt{n_{\mathrm{meas}, \nu} }
        \sqrt{n_{\mathrm{meas}, \nu'} }
    \cdot
    \nonumber \\
    \cdot
        \sum_{\ell = 0}^{n_{\mathrm{ch}} - 1 }
        \int_{-\infty}^\infty
        dE
        \rho_{h \ell} (E - \varepsilon_{h \ell})
        \frac{
            v^{(\nu)}_{h \ell}
            v^{(\nu')}_{h \ell}
        }{z + E }
    \label{avr_spectra_using_QPE:gf_rec_in_NO_using_DOS_h}
\end{gather}
for the hole part.
$n_{\mathrm{meas}, \nu}$ is the occupancy of the $\nu$th NO,
constructed by diagonalizing the measured one-electron matrix elements $\gamma_{\mathrm{meas}}.$
$\{ \varepsilon_{\xi \ell} \}_{\ell = 0}^{n_{\mathrm{ch}} - 1}$ are fictitious $\xi$ excitation energies.
$n_{\mathrm{ch}}$ is the number of fictitious excitation channels, which is not necessarily equal to that of the true excitation channels.
We require $n_{\mathrm{sorb}} \leq n_{\mathrm{ch}}.$
$v_{\xi \ell}^{(\nu)}$ are fictitious transition amplitudes in the $\ell$th channel.
They correspond to the true transition amplitudes for the NOs in 
Eq.~(\ref{avr_spectra_using_QPE:def_transition_vec_for_NO}).
$\rho_{\xi \ell}$ is fictitious DOS,
introduced for capturing the overall features of crowded peaks on the energy axis.
We require $\rho_{\xi \ell}$ to be centered at the origin and its integral to be normalized to unity.
We adopt the quadratic DOS (see Appendix \ref{sec:gf_from_fict_dos}) and
their widths $\{ \Delta E_{\xi \ell} \}_{\xi, \ell}$ as trial parameters.
One can also locate an isolated peak by using a delta function instead of the fictitious DOS.
Summarizing, we adopt
\begin{align}
    \Lambda_\xi
    \equiv
    \left\{
        \varepsilon_{\xi \ell},
        \Delta E_{\xi \ell},
        v_{\xi \ell}^{(\nu)}
    \right\}_{\ell, \nu}
\end{align}
as the trial parameters for $\xi$ excitations to be optimized in QAVG by referring to the data 
$\{ f^{(p, m)}_{\xi j}, f^{(p, m m')}_{\xi \sigma j} \}_{p, m, m', \sigma, j}.$

Since the true transition amplitudes for NOs satisfy the orthonormalization conditions in
Eq.~(\ref{avr_spectra_using_QPE:sum_nat_orb_trans_vec}),
similar conditions should be introduced for restricting the search space to reach physically plausible fictitious transition amplitudes.  
We therefore employ the hyperspherical Householder parametrization (see Appendix \ref{sec:Householder_for_orthogonal_vectors}) to express
$\{ v_{\xi \ell}^{(\nu)} \}_{\ell, \nu}$
as $n_{\mathrm{sorb}}$ vectors in $\mathbb{R}^{n_{\mathrm{ch}}},$
so that
\begin{align}
    \boldsymbol{v}_{\xi}^{(\nu)} \cdot \boldsymbol{v}_{\xi}^{(\nu')}
    =
        \delta_{\nu \nu'}
    \label{avr_spectra_using_QPE:orthonormality_btwn_fict_trans_vecs}
\end{align}
holds by construction.

\subsubsection{Modeling probability distributions for QPE}

In order for the trial parameters introduced just above to be located in the framework of QAVG,
we need to model the probability distributions.
In other words, we need to {\it define} probability distributions
for the QPE measurement results that would be obtained when the GF of an input state was
$\widetilde{G}_{\mathrm{rec}} (z; \Lambda)$ in
Eqs.~(\ref{avr_spectra_using_QPE:gf_rec_in_NO_using_DOS_e}) and
(\ref{avr_spectra_using_QPE:gf_rec_in_NO_using_DOS_h}).
To this end,
by referring to the relation between the probability distributions with the true GF derived in Appendix \ref{sec:exc_energy_meas_for_Gibbs},
we define the reasonable probability distributions for a fictitious input state characterized by the trial parameters.
Their expressions are provided in Appendix \ref{sec:modeled_prob_distrs}.

\section{Results and discussion}

\subsection{Computational details}

For the DFT calculations,
we used Quantum Espresso 
\cite{Quantum_Espresso_refs_1,Quantum_Espresso_refs_2,Quantum_Espresso_refs_3} with the ultrasoft pseudopotentials \cite{bib:46} for plane-wave basis sets and the Perdew--Burke--Ernzerhof (PBE) \cite{bib:37} functional for exchange correlation energies.
We performed the DFT calculations for SrVO$_3$ in a cubic perovskite structure with its lattice constant $3.841$\AA \cite{bib:3158}. 
We sampled $N_{\boldsymbol{k}} = 6 \times 6 \times 6$ $k$ points for the self-consistent field calculations,
from which we constructed the MLWOs by using Wannier90 \cite{Wannier90}.
Since the electronic structure near the Fermi level of SrVO$_3$ crystal consists of the $t_{2 g}$ band manifold of vanadium ions comprising the VO$_6$ octahedra,
we constructed $n_{\mathrm{corr}} = 3$ MLWOs per spin for the AIM.

For the FCI-based DMFT calculations,
we used a modified version of DCore \cite{bib:7027} and a homemade FCI solver based on the Arnoldi method \cite{bib:ARPACK}.
We adopted the Kanamori interaction parameters ($U = 3.44$ eV,
$U_0 = 2.49$ eV, $J = 0.46$ eV) in the second-quantized Hamiltonian $\mathcal{H}$ for the AIM of SrVO$_3$ \cite{bib:7218, bib:7027}.
We used $n_{\mathrm{bath}} = 3$ bath sites for the impurity problem involving $n_{\mathrm{sorb}} = 12$ spin orbitals.
The exact GFs were calculated by employing the Lanczos method \cite{bib:3064} from the FCI results with the temperature $\beta^{-1} \equiv 0.025$ eV.
We took only the excitations into account where the Boltzmann factors $e^{-\beta E_{\lambda_0} }$ [see Eq.~(\ref{G_imag_freq_sum_partial_G})] are larger than the threshold $\varepsilon_{\mathrm{B}} \equiv 10^{-4}.$
Although circuit simulations for $\mathcal{C}_{\mathrm{GF-QPE}}$ can provide QPE histograms for a finite number of measurements,
we adopted the exact probability distributions instead of them based on the FCI results.
It was because we would like to focus on the validity of the reconstruction of GF using the trial parameters without being worried about the statistical errors.
We adopted $n_{\mathrm{Matsu}} = 2047$ for the Matsubara frequencies $\omega_n \ (n = 0, \pm 1, \dots, \pm n_{\mathrm{Matsu}} ).$

In the QAVG calculations, 
We optimized the trial parameters based on Monte Carlo method,
where we varied $\tau_{\mathrm{Metro}}$ in the factor $\exp(-\tau_{\mathrm{Metro}} F (\Lambda) )$ for the Metropolis sampling.
Specifically, we began with $\tau_{\mathrm{Metro}} < 1000$ and increased it linearly as the Monte Carlo steps proceed until $\tau_{\mathrm{Metro}} = 4000.$
We performed 40000 steps at each Monte Carlo run.
At each DMFT iteration, we carried out 256 runs, from which we adopted the trial values that achieved the lowest cost among the runs as the optimal ones.
The trial parameters were updated during the steps basically by changing their values randomly within their own specified ranges.
In addition, the values of parameters were exchanged at a certain probability.

\subsection{DMFT}

\subsubsection{Excitation channels}

\begin{figure}
\begin{center}
\includegraphics[width=6cm]{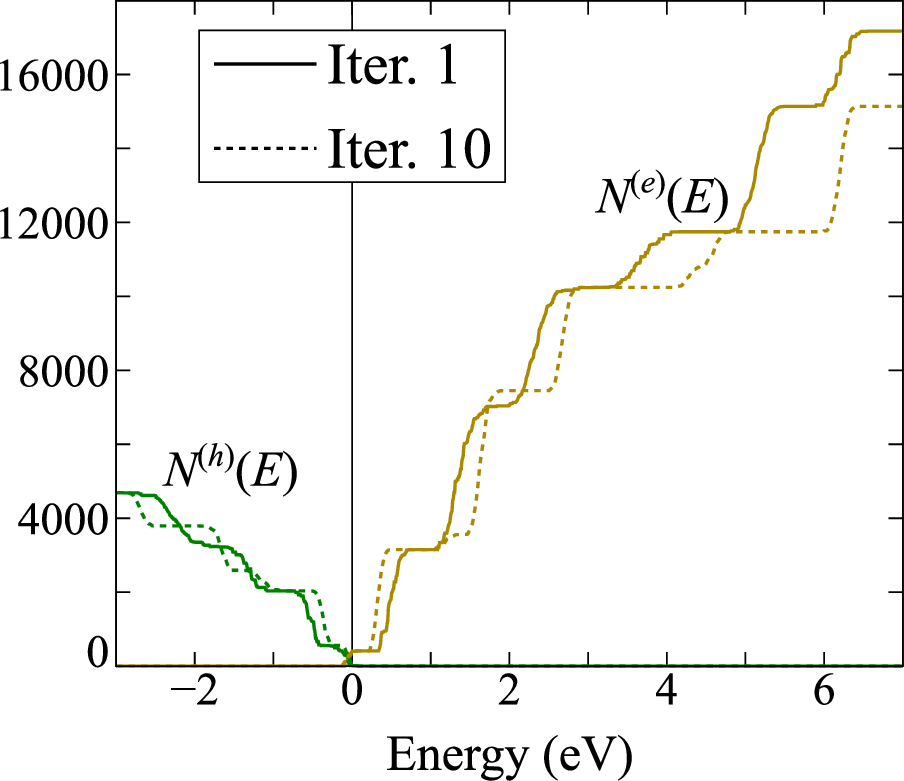}
\end{center}
\caption{
Accumulated numbers of excitation channels for electron and hole excitations as functions of energy $E$ at the nonzero temperature in the AIM for FCI-DMFT.
Those at the first and tenth iterations are plotted.
They represent the excitations that contribute to the GF with the Boltzmann factors larger than $\varepsilon_{\mathrm{B}}.$
}
\label{fig:num_of_excitations_in_gf}
\end{figure}

As described above, the QAVG approach in the present study uses fictitious physical quantities instead of the genuine physical quantities.
We confirm here that this approach is suitable for the present system being considered. 
To this end, we define the accumulated number of electron excitation channels as
\begin{align}
    N^{(e)} (E)
    \equiv
        \sum_{\exp (-\beta E_{\lambda_0}) > \varepsilon_{\mathrm{B}} }
        \sum_{\lambda}
        \theta (E - (E_\lambda - E_{\lambda_0} ))
    ,
    \label{avr_spectra_using_QPE:def_num_exc_e}
\end{align}
where the summation is over the electron-excited states.
$\theta$ is the step function.
We define the accumulated number of hole excitation channels similarly as
\begin{align}
    N^{(h)} (E)
    \equiv
        \sum_{\exp (-\beta E_{\lambda_0}) > \varepsilon_{\mathrm{B}} }
        \sum_{\lambda}
        \theta (-E - (E_\lambda - E_{\lambda_0} ))
    ,
    \label{avr_spectra_using_QPE:def_num_exc_h}
\end{align}
where the summation is over the hole-excited states.
The accumulated numbers of excitations in the AIM for the first and tenth iterations are plotted in Fig.~\ref{fig:num_of_excitations_in_gf}.
We see that thousands of channels may contribute to each cluster of spectral peaks due to the nonzero temperature.
This fact implies that the adoption of fictitious parameters whose numbers are much smaller than the genuine physical quantities is a practically suitable way for reproducing observed histograms at a nonzero temperature.

\subsubsection{One-shot QAVG after FCI-DMFT}

\begin{figure*}
\begin{center}
\includegraphics[width=16cm]{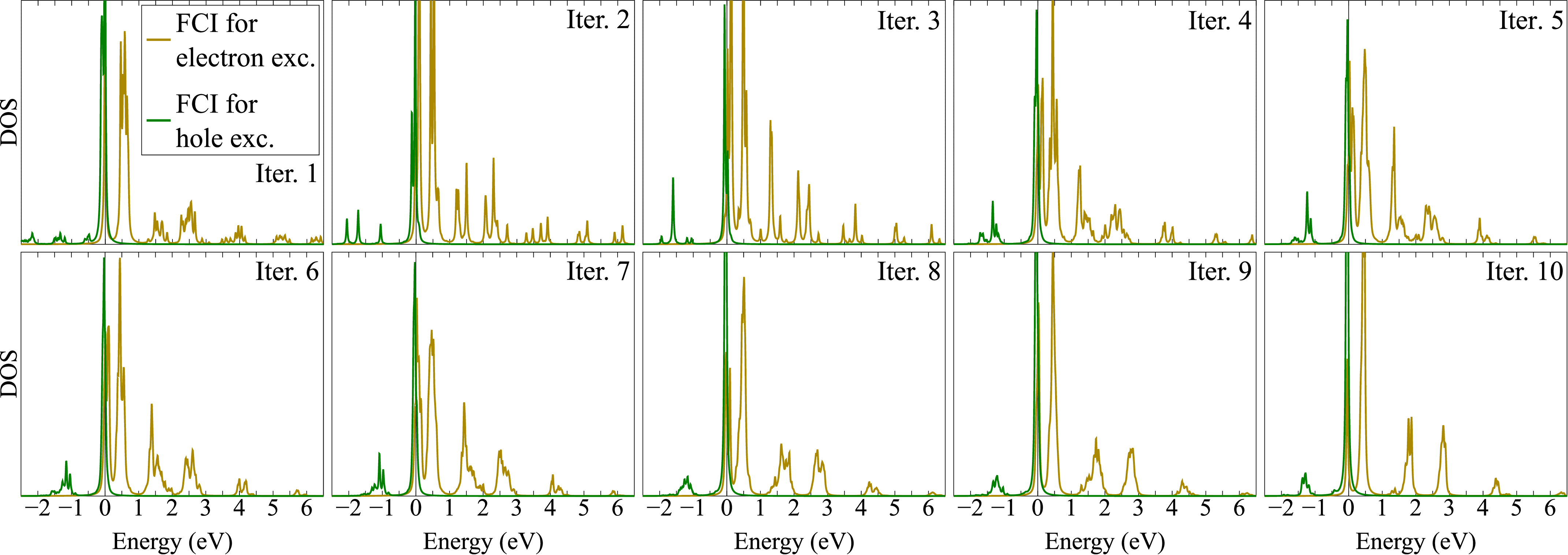}
\end{center}
\caption{
FCI-DOS of the AIM at each iteration of FCI-DMFT.
}
\label{fig:fci_dos}
\end{figure*}

Figure \ref{fig:fci_dos} shows the FCI-GF of the AIM at each iteration during the FCI-DMFT loop.
As expected in the discussion above,
many excitation channels enter the spectra due to the finite temperature.
In addition, we see the clusters, each of which consists of crowded spectral peaks, at late iterations.
These features in the present system are favorable for reconstructing by introducing much fewer trial parameters than the excitation channels.

\begin{figure*}
\begin{center}
\includegraphics[width=14cm]{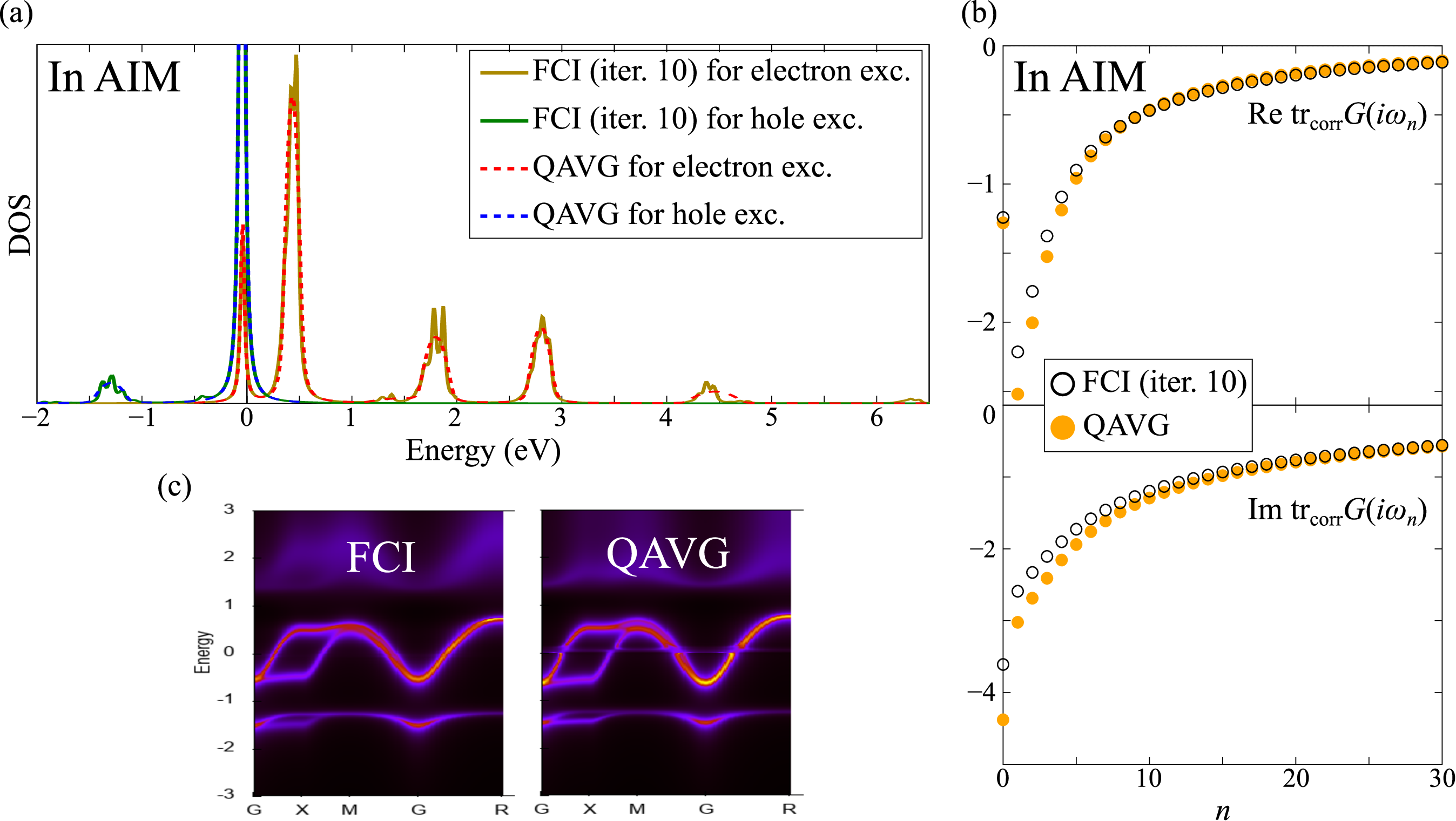}
\end{center}
\caption{
(a)
One-shot QAVG-DOS of the AIM that reconstructs the FCI-DOS at the tenth iteration of FCI-DMFT.
(b)
Traces of GFs at the Matsubara frequencies over the correlated orbitals.
(c)
Momentum-resolved DOS in SrVO$_3$ from the FCI-DMFT and the one-shot QAVG-DMFT results.
}
\label{fig:one_shot_dos_iter10}
\end{figure*}

To see how the central idea of QAGV works,
we reconstructed the FCI-GF of the AIM at the tenth iteration.
Specifically, we performed the QPE simulations for $n_{q \mathrm{val}} = 7$ ancillary qubits by using the origin shift $\Delta \equiv s/(3 t_0) \ (s = 0, 1, 2)$ with the grid spacing $1/t_0 = 1/6$ eV.
We adopted the probability distributions for these QPE circuits as the input histograms to QAVG calculations.
We used the decay rate $\tau_{\mathrm{dec}} = 1$ eV$^{-1}$ in the nonuniform-weight $L_1$ distance.
We introduced 6 (2) independent fictitious excitation energies with each at most sixfold degenerate for the electron (hole) part.
Figure \ref{fig:one_shot_dos_iter10}(a) shows the FCI-DOS at the tenth iteration and the reconstructed QAVG-DOS.
We see that the overall features of the FCI-DOS are well captured by the QAVG-DOS despite the small number of trial parameters.
To examine the degree of the agreement between the FCI- and QAVG-DOS more in detail,
we plot the traces of GFs at the Matsubara frequencies over the correlated orbitals in \ref{fig:one_shot_dos_iter10}(b).
It is found that there exist discrepancies between the GF from FCI and that from QAVG mainly in the low-energy (small $n$) regime.
We calculated the momentum-resolved DOS in the original periodic system based on the self-energies from the FCI and QAVG-DMFT results,
as shown in Fig.~\ref{fig:one_shot_dos_iter10}(c).
Although small discrepancies of spectra are seen in the low-energy regime in the electron excitations,
the overall features of the FCI spectra are well reproduced by the QAVG spectra.
We confirmed that the ordinary $L_1$ distance $(\tau_{\mathrm{dec}} = 0)$ leads to worse reproduction of the FCI spectra (not shown).

\subsubsection{Iterative QAVG-DMFT}

Here we examine the iterative QAVG-DMFT, that is, the GF of the AIM is reconstructed at each iteration as illustrated in Fig.~\ref{fig:dft_dmft_qavg_flowchart}.
In the QAVG calculations, we introduced at most 8 and 4 independent fictitious excitation energies for the electron and hole parts, respectively.
We assumed each of the independent fictitious excitation energies to be threefold degenerate,
so that the numbers of hyperspherical parameters are at most
123 and 51 for the electron and hole parts, respectively [see Eq.~(\ref{Householder_for_orthogonal_vectors:num_of_params})], for the $n_{\mathrm{sorb}}/2$ spin orbitals per spin.

\begin{figure}
\begin{center}
\includegraphics[width=6cm]{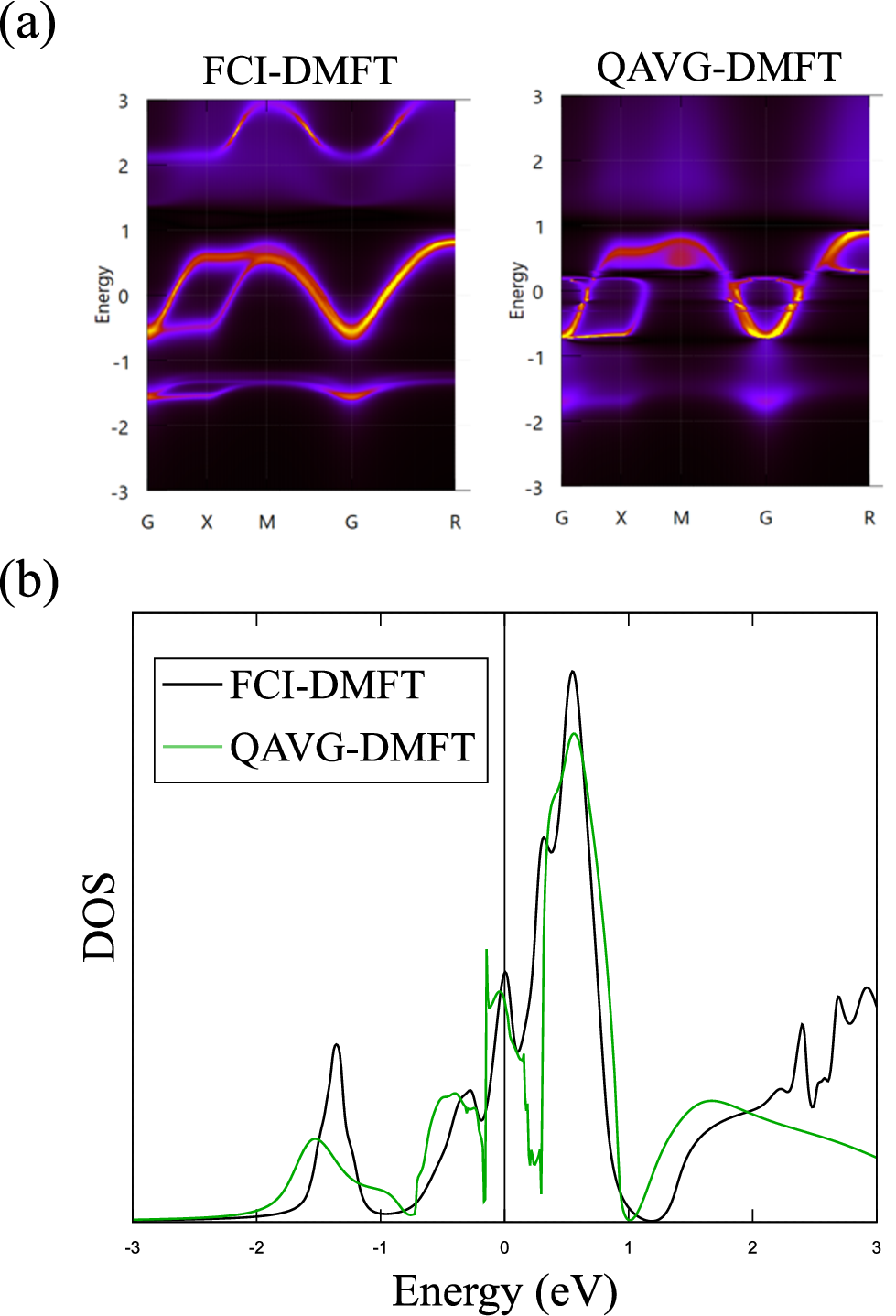}
\end{center}
\caption{
(a)
Momentum-resolved DOS in SrVO$_3$ from FCI-DMFT and QAVG-DMFT.
(b)
DOS in SrVO$_3$ from FCI-DMFT and QAVG-DMFT.
}
\label{fig:tentative_dos_fci_and_qavg}
\end{figure}

We calculated the momentum-resolved DOS in the original periodic system,
as shown in Fig.~\ref{fig:tentative_dos_fci_and_qavg}(a).
The momentum-integrated DOS are also shown in (b) of the figure.
While the overall features of the FCI-DOS are reasonably reproduced by the QAVG-DOS,
the latter exhibits dips near $\pm 0.2$ eV away the origin.

\section{Conclusions}

We proposed the quantum-classical hybrid scheme for performing a DMFT calculation at a finite temperature based on the QAVG approach.
The quantum part of the scheme uses the modified QPE circuits for sampling the one-particle GF without knowing the excitation channel invoked at each measurement.
Those circuits render the sampling process efficient as long as the input Gibbs state has been prepared since there is no need for numerical diagonalization of the correlated subspace.
The classical part of the scheme, for which the QAVG approach is responsible,
introduces trial parameters whose numbers are far less than those of genuine physical quantities involved in the excitations.
By modeling the probability distributions via the trial parameters,
the QAVG approach finds the optimal ones with alleviating the biases coming from the finite resolution of QPE grids.
We applied the QAVG-DMFT scheme to SrVO$_3$ to demonstrated its validity via numerical simulations.

Since the QAVG approach itself does not assume specific electronic-structure methodologies,
it can work independently for obtaining the GF of a correlated system.
We will report the application of this approach to a correlated system by using real quantum computers elsewhere.

\begin{acknowledgments}
The authors are grateful to Dr.~Hiroshi Shinaoka for fruitful discussion.
This work was partially supported by the Center of Innovations for Sustainable Quantum AI (JST Grant Number JPMJPF2221).
The computation in this work has been done using the facilities of the Supercomputer Center, the Institute for Solid State Physics, the University of Tokyo (ISSPkyodo-SC-2026-Ea-0014).
\end{acknowledgments}

\begin{widetext}

\appendix

\section{Construction of excitation circuits}
\label{sec:excitation_circuits}

Here we briefly review the implementation of excitation circuits for the one-particle GF of a many-electron system,
where the electron number increases or decrease by one \cite{bib:5005}.

Let us consider a set of orthonormalized WBOs.
For the $m$th WBO,
we define the operators $U_{0 m} \equiv a_m + a_m^\dagger$ and $U_{1 m} \equiv a_m - a_m^\dagger.$
They are unitary due to the fermionic anti-commutation relation $\{ a_m, a_{m'} \} = \delta_{m m'}.$
This means that they are implementable as unitary gates regardless of the adopted mapping scheme to qubit representation.

\subsection{Diagonal excitation circuits}

The diagonal excitation circuit $\mathcal{C}_m$ for the $m$th WBO is shown in 
Fig.~\ref{fig:circuit_excitation}(a), equipped with a single ancilla $(n_{q \mathrm{exc}} = 1).$
One can confirm that an arbitrary input state $| \Psi_{\mathrm{in}} \rangle$
undergoes the unitary gates and the state of the entire system is
\begin{align}
    a_m | \Psi_{\mathrm{in} } \rangle| 0 \rangle
    +
    a_m^\dagger | \Psi_{\mathrm{in} } \rangle| 1 \rangle       
    \label{avr_spectra_using_QPE:states_before_meas_in_diag}
\end{align}
immediately before the measurement.

\subsection{Off-diagonal excitation circuits}

For the pair of $m$th and $m'$th WBOs $(m \ne m'),$
we define the following four auxiliary operators \cite{bib:5005}:
\begin{gather}
    a_{m m'}^{\pm}
    \equiv
        \frac{
            a_m
            \pm
            e^{-i \pi/4}
            a_{m'}
        }{2}
    , \
    a_{m m'}^{\pm \dagger}
    \equiv
        \frac{
            a_m^\dagger
            \pm
            e^{i \pi/4}
            a_{m'}^\dagger
        }{2}
    .
    \label{avr_spectra_using_QPE:def_aux_c_and_a_oprs}
\end{gather}
The off-diagonal excitation circuit $\mathcal{C}_{m m'}$ is shown in 
Fig.~\ref{fig:circuit_excitation}(b), equipped with two ancillae $(n_{q \mathrm{exc}} = 2).$
One can confirm that an arbitrary input state $| \Psi_{\mathrm{in}} \rangle$
undergoes the unitary gates and the state of the entire system is
\begin{gather}
        e^{i \pi/4} a_{m' m}^+
        | \Psi_{\mathrm{in}} \rangle
        \overbrace{| 0 \rangle}^{| q_{\mathrm{A1}} \rangle}
        \overbrace{| 0 \rangle}^{| q_{\mathrm{A0}} \rangle}
        +
        a_{m m'}^{+ \dagger} 
        | \Psi_{\mathrm{in}} \rangle
        | 0 \rangle | 1 \rangle
        -
        e^{i \pi/4} a_{m' m}^- 
        | \Psi_{\mathrm{in}} \rangle
        | 1 \rangle | 0 \rangle
        +
        a_{m m'}^{- \dagger}
        | \Psi_{\mathrm{in}} \rangle
        | 1 \rangle | 1 \rangle
    \label{avr_spectra_using_QPE:states_before_meas_in_off_diag}
\end{gather}
immediately before the measurement.

\begin{figure}
\begin{center}
\includegraphics[width=16cm]{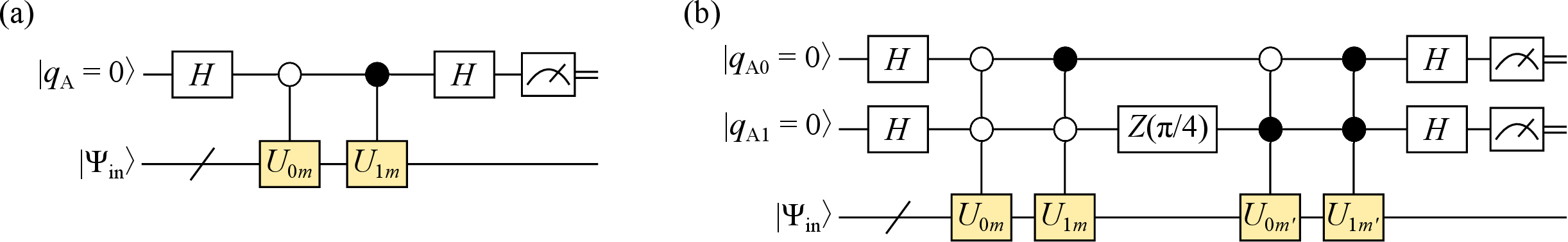}
\end{center}
\caption{
(a)
Diagonal excitation circuit $\mathcal{C}_m$ for the one-particle GF in terms of the $m$th WBO.
$H$ in the figure represents the Hadamard gate.
(b)
Off-diagonal excitation circuit $\mathcal{C}_{m m'}$ for the one-particle GF in terms of the $m$th and $m'$th WBOs.
These circuits were proposed in Ref.~\cite{bib:5005}.
}
\label{fig:circuit_excitation}
\end{figure}

\section{Action of excitation circuits on Gibbs state}
\label{sec:excitation_on_gibbs_state}

Here we analyze the action of the excitation circuits described in 
Appendix \ref{sec:excitation_circuits} on the Gibbs state.

\subsection{Gibbs state as an input}

The density operator of the Gibbs state at an inverse temperature $\beta$ is given by
$
\rho_{\mathrm{Gibbs}}
=
e^{-\beta \mathcal{H}}/\mathcal{Z}
=
\sum_{\lambda}
e^{-\beta E_{\lambda}}
| \Psi_{\lambda} \rangle \langle \Psi_{\lambda} |/\mathcal{Z}
.
$
We want to derive the expression of the final state when 
$\rho_{\mathrm{Gibbs}}$ is input to the channel-agnostic circuit $\mathcal{C}_{\mathrm{GF-QPE}}$
in Fig.~\ref{fig:circuit_gf_for_gibbs}(b).
Since the Gibbs state is a classical mixture of the energy eigenstate,
we can find the final state by tracing the constituent energy eigenstates separately and summing up them according to the Boltzmann weights.
Furthermore, the controlled RTE operations before the excitation part in $\mathcal{C}_{\mathrm{GF-QPE}}$ act only as giving a phase factor to an input energy eigenstate, that is, the physical state is unaffected.
This means that we can examine the probability distributions of the measurement outcomes from the excitation circuits without paying attention to the phase factor.

\subsection{Diagonal excitation}

For the case of the diagonal circuit $\mathcal{C}_m,$
the probability that a hole excitation occurs on the Gibbs state is nothing but the probability that the ancilla is observed to be $| 0 \rangle$ [see Eq.~(\ref{avr_spectra_using_QPE:states_before_meas_in_diag})].
The probability is thus
\begin{align}
    \mathbb{P}^{(m)}_h
    =
        \frac{1}{\mathcal{Z}}
        \sum_{\lambda_0}
        e^{-\beta E_{\lambda_0} }
        \langle \Psi_{\lambda_0} |
        a_m^\dagger
        a_m
        | \Psi_{\lambda_0} \rangle
    =
        \langle a_m^\dagger a_m \rangle
    ,
    \label{avr_spectra_using_QPE:prob_distr_for_diag_el}
\end{align}
where the brakets sandwiching the operators indicate the thermal average.
This is equal to the occupancy of the $m$th WBO, that is,
the diagonal component of the one-electron matrix element
\begin{align}
    \gamma_{m m'}
    \equiv
        \langle a_m^\dagger a_{m'} \rangle
    .
    \label{avr_spectra_using_QPE:def_one_electron_mat_el}
\end{align}
The matrix $\gamma$ is clearly Hermitian: $\gamma_{m' m} = \gamma_{m m'}^*.$
The probability of an electron excitation is
$\mathbb{P}^{(m)}_e = 1 - \mathbb{P}^{(m)}_h.$

Let us analyze the statistical error for $M_{\mathrm{d}}$ shots of measurements on the ancilla of the $\mathcal{C}_m$ circuit.
The number $M^{(m)}_h$ of outcomes $| 0 \rangle$ obeys the binomial distribution.
The estimated diagonal matrix element
\begin{align}
    \gamma_{\mathrm{meas}, m m}
    \equiv
        \frac{M^{(m)}_h}{M_{\mathrm{d}}}
\label{avr_spectra_using_QPE_ruin:one_el_mat_diag_from_meas}
\end{align}
thus has the variance of
\begin{align}
    \mathrm{Var} [\gamma_{\mathrm{meas}, m m}]
    =
        \frac{1}{M_{\mathrm{d}}^2}
        \mathrm{Var} \left[ M^{(m)}_h \right]
    =
        \frac{\mathbb{P}^{(m)}_h (1 - \mathbb{P}^{(m)}_h)}{M_{\mathrm{d}}}
        .
\end{align}
Since $x (1 - x) \leq 1/4$ for any $x$ satisfying $0 \leq x \leq 1,$
the upper bound of the variance is given by
\begin{align}
    \mathrm{Var} [\gamma_{\mathrm{meas}, m m}]
    \leq
        \frac{1}{4 M_{\mathrm{d}} }
    .
\end{align}
The probability that the difference between the estimated occupancy of the $m$th WBO and the exact value is larger than or equal to $\epsilon$ is thus upper bounded by the Chebyshev's inequality as
\begin{align}
    \mathrm{Prob}
    \left(
        \left| \gamma_{\mathrm{meas}, m m} - \gamma_{m m}
        \right|
        \geq
            \epsilon
    \right)
    \leq
        \frac{1}{\epsilon^2}
        \mathrm{Var}
        \left[ \gamma_{\mathrm{meas}, m m} \right]
    \leq
        \frac{1}{4 M_{\mathrm{d}} \epsilon^2}
    .
\end{align}
This tells us that we can achieve an accuracy $\epsilon$ with a failure probability $p_{\mathrm{fail}}$ if the number of shots satisfies
\begin{align}
    M_{\mathrm{d}}
    \geq
        \frac{1}{4 \epsilon^2 p_{\mathrm{fail}} }
    .
    \label{avr_spectra_using_QPE:num_of_required_shots_diag}
\end{align}

\subsection{Off-diagonal excitation}

For the case of the off-diagonal circuit $\mathcal{C}_{m m'},$
the probabilities
$\mathbb{P}^{(m m')}_{\xi \sigma } \ (\xi \in \{e, h\}, \ \sigma \in \{ +, - \})$
of the four possible measurement outcomes
$| q_{\mathrm{A1}} \rangle | q_{\mathrm{A0}} \rangle$
are calculated from
Eq.~(\ref{avr_spectra_using_QPE:states_before_meas_in_off_diag}) as
\begin{align}
\begin{split}
    | 0 \rangle | 0 \rangle: \
    \mathbb{P}^{(m m')}_{h +}
    &\equiv
        \langle 
        a_{m' m}^{+ \dagger}
        a_{m' m}^+
        \rangle
    =
        \frac{
            \gamma_{m m}
            +
            \gamma_{m' m'}
        }{4}
        +
        \frac{1}{2}
        \mathrm{Re}
        \left( e^{i \pi/4} \gamma_{m m'} \right)
    ,
    \\
    | 0 \rangle | 1 \rangle: \
    \mathbb{P}^{(m m')}_{e +}
    &\equiv
        \langle
        a_{m m'}^+
        a_{m m'}^{+ \dagger}
        \rangle
    =
        \frac{1}{2}
        -
        \frac{
            \gamma_{m m}
            +
            \gamma_{m' m'}
        }{4}
        -
        \frac{1}{2}
        \mathrm{Re}
        \left( e^{-i \pi/4} \gamma_{m m'} \right)
    \\
    | 1 \rangle | 0 \rangle: \
    \mathbb{P}^{(m m')}_{h -}
    &\equiv
        \langle
        a_{m' m}^{- \dagger}
        a_{m' m}^-
        \rangle
    =
        \frac{
            \gamma_{m m}
            +
            \gamma_{m' m'}
        }{4}
        -
        \frac{1}{2}
        \mathrm{Re}
        \left( e^{i \pi/4} \gamma_{m m'} \right)
    \\
    | 1 \rangle | 1 \rangle: \
    \mathbb{P}^{(m m')}_{e -}
    &\equiv
        \langle
        a_{m m'}^-
        a_{m m'}^{- \dagger}
        \rangle
    =
        \frac{1}{2}
        -
        \frac{
            \gamma_{m m}
            +
            \gamma_{m' m'}
        }{4}
        +
        \frac{1}{2}
        \mathrm{Re}
        \left( e^{-i \pi/4} \gamma_{m m'} \right)
    .
\end{split}
\label{avr_spectra_using_QPE:prob_distr_for_off_diag_excitation}
\end{align}
It is noted that 
$\mathbb{P}^{(m m')}_{\xi \sigma}$ is not equal to $\mathbb{P}^{(m' m)}_{\xi \sigma}$ in general.
The off-diagonal component of the one-electron matrix element can be written as
\begin{align}
\begin{split}
    \mathrm{Re} \gamma_{m m'}
    &=
        -
        \frac{1}{\sqrt{2}}
        +
        \sqrt{2}
        \left(
            \mathbb{P}^{(m m')}_{h +}
            +
            \mathbb{P}^{(m m')}_{e -}
        \right)
    ,
    \\
    \mathrm{Im} \gamma_{m m'}
    &=
        \frac{1}{\sqrt{2}}
        -
        \sqrt{2}
        \left(
            \mathbb{P}^{(m m')}_{h +}
            +
            \mathbb{P}^{(m m')}_{e +}
        \right)
    .
\end{split}
\label{avr_spectra_using_QPE:one_el_mat_from_prob_m_mp}
\end{align}
Since $\gamma$ is Hermitian,
we can estimate $\gamma_{m m'}$ and $\gamma_{m' m}$ from repeated measurements using only the $\mathcal{C}_{m m '}$ circuit.
As for estimating the off-diagonal components of the GF, however,
we need to use not only $\mathcal{C}_{m m'}$ but also $\mathcal{C}_{m' m}.$

Let us analyze the statistical error for $M_{\mathrm{od}}$ shots of measurements on the ancillae of the $\mathcal{C}_{m m'}$ circuit.
The four numbers 
$M^{(m m')}_{h+}, M^{(m m')}_{e+}, M^{(m m')}_{h-},$ and $M^{(m m')}_{e-}$
of respective outcomes in
Eq.~(\ref{avr_spectra_using_QPE:prob_distr_for_off_diag_excitation})
obey the multinomial distribution specified by
$\{ \mathbb{P}^{(m m')}_{\xi \sigma} \}_{\xi, \sigma}.$
Since the exact value of the off-diagonal component of the one-electron matrix element satisfies Eq.~(\ref{avr_spectra_using_QPE:one_el_mat_from_prob_m_mp}),
we can estimate its real and imaginary parts by collecting the results of $M_{\mathrm{od}}$ shots as
\begin{align}
\begin{split}
    \mathrm{Re} \gamma_{\mathrm{meas}, m m'}
    &\equiv
            -
            \frac{1}{\sqrt{2}}
            +
            \sqrt{2}
            \left(
                \frac{M^{(m m')}_{h+} }{M_{\mathrm{od}} }
                +
                \frac{M^{(m m')}_{e-} }{M_{\mathrm{od}} }
            \right)
    ,
    \\
    \mathrm{Im} \gamma_{\mathrm{meas}, m m'}
    &\equiv
            \frac{1}{\sqrt{2}}
            -
            \sqrt{2}
            \left(
                \frac{M^{(m m')}_{h+} }{M_{\mathrm{od}} }
                +
                \frac{M^{(m m')}_{e+} }{M_{\mathrm{od}} }
            \right)
        .
\end{split}
\label{avr_spectra_using_QPE:one_el_mat_off_diag_from_meas}
\end{align}
Their variances are calculated as
\begin{align}
\begin{split}
    \mathrm{Var}
    \left[ \mathrm{Re} \gamma_{\mathrm{meas}, m m'} \right]
    &=
        \frac{2}{M_{\mathrm{od}}^2 }
        \mathrm{Var}
        \left[ M^{(m m')}_{h+} + M^{(m m')}_{e-} \right]
    =
        \frac{2}{M_{\mathrm{od}} }
            \left(
                \mathbb{P}^{(m m')}_{h +}
                +
                \mathbb{P}^{(m m')}_{e -}
            \right)
            \left(
                1
                -
                \left(
                    \mathbb{P}^{(m m')}_{h +}
                    +
                    \mathbb{P}^{(m m')}_{e -}
                \right)
            \right)
    ,
    \\
    \mathrm{Var}
    \left[ \mathrm{Im} \gamma_{\mathrm{meas}, m m'} \right]
    &=
        \frac{2}{M_{\mathrm{od}}^2 }
        \mathrm{Var}
        \left[ M^{(m m')}_{h+} + M^{(m m')}_{e+} \right]
    =
        \frac{2}{M_{\mathrm{od}} }
            \left(
                \mathbb{P}^{(m m')}_{h +}
                +
                \mathbb{P}^{(m m')}_{e +}
            \right)
            \left(
                1
                -
                \left(
                    \mathbb{P}^{(m m')}_{h +}
                    +
                    \mathbb{P}^{(m m')}_{e +}
                \right)
            \right)
    .
\end{split}
\end{align}
The upper bounds of these variances are thus given by
\begin{align}
    \mathrm{Var}
    \left[ \mathrm{Re} \gamma_{\mathrm{meas}, m m'} \right]
    \leq
        \frac{1}{2 M_{\mathrm{od}} }
    , \
    \mathrm{Var}
    \left[ \mathrm{Im} \gamma_{\mathrm{meas}, m m'} \right]
    \leq
        \frac{1}{2 M_{\mathrm{od}} }
        .
\end{align}
The probability that the difference between the estimated real part and the exact value is larger than or equal to $\epsilon$ is thus upper bounded by the Chebyshev's inequality as
\begin{align}
    \mathrm{Prob}
    \left(
        \left|
            \mathrm{Re} \gamma_{\mathrm{meas}, m m'}
            -
            \mathrm{Re} \gamma_{m m}
        \right|
        \geq
            \epsilon
    \right)
    \leq
        \frac{1}{\epsilon^2}
        \mathrm{Var}
        \left[ \mathrm{Re} \gamma_{\mathrm{meas}, m m} \right]
    \leq
        \frac{1}{2 M_{\mathrm{od}} \epsilon^2}
    .
\end{align}
This means that We can achieve an accuracy $\epsilon$ with a failure probability $p_{\mathrm{fail}}$ for the real part of $\gamma_{\mathrm{meas}, m m'}$ if the number of shots satisfies
\begin{align}
    M_{\mathrm{od}}
    \geq
        \frac{1}{2 \epsilon^2 p_{\mathrm{fail}} }
    .
    \label{avr_spectra_using_QPE:num_of_required_shots_off_diag}
\end{align}
We can derive a similar condition for the imaginary part.

From Eqs.~(\ref{avr_spectra_using_QPE:num_of_required_shots_diag})
and (\ref{avr_spectra_using_QPE:num_of_required_shots_off_diag}),
the total number $M_{\mathrm{tot}}$ of shots for determining all the components of $\gamma$ with the specified accuracy and failure probability needs to satisfy
\begin{align}
    M_{\mathrm{tot}}
    =
        n_{\mathrm{sorb}}
        M_{\mathrm{d}}
        +
        (n_{\mathrm{sorb}}^2 - n_{\mathrm{sorb}})
        M_{\mathrm{od}}
    \geq
        \frac{1}{\epsilon^2 p_{\mathrm{fail}} }
        \left(
            \frac{n_{\mathrm{sorb}}^2 }{2}
            -
            \frac{n_{\mathrm{sorb}} }{4}
        \right)
    .
\end{align}

\section{Excitation-energy measurement for Gibbs state}
\label{sec:exc_energy_meas_for_Gibbs}

\subsection{Diagonal excitation and QPE}

Let us assume that the number $n_{q \mathrm{val}}$ of the ancillae in the QPE circuit for binary representation of excitation energies is sufficiently large.
We already know the action of $\mathcal{C}_{\mathrm{GF-QPE}}$ on an energy eigenstate,
as provided in Eq.~(\ref{avr_spectra_using_QPE:circ_GF_QPE_for_energy_eigenstate}).
When the Gibbs state is input to $\mathcal{C}_{\mathrm{GF-QPE}}$ involving $\mathcal{C}_m$ for the $m$th WBO,
the probability that $\xi$ excitation occurs and an excitation energy $\varepsilon$ in binary representation is observed is calculated as
\begin{align}
    \mathbb{P}^{(m)}_{\xi} (\varepsilon)
    =
        \frac{1}{\mathcal{Z}}
        \sum_{\lambda_0, \lambda}
        e^{-\beta E_{\lambda_0} }
        | \langle \Psi_\lambda | a_m^{\dagger} | \Psi_{\lambda_0 } \rangle |^2
        \delta_{E_{\lambda} - E_{\lambda_0}, \varepsilon }
    =
        S^{(\xi)}_{m m} (\varepsilon)
    ,
\label{avr_spectra_using_QPE:prob_distr_with_exc_energy_for_diag_QPE}
\end{align}
where we used the definition of the spectral matrix in
Eq.~(\ref{avr_spectra_using_QPE:def_spectral_mat})
to get the second equality.
This indicates that
Eq.~(\ref{avr_spectra_using_QPE:prob_distr_with_exc_energy_for_diag_QPE})
relates the probability distribution of measurements in $\mathcal{C}_{\mathrm{GF-QPE}}$ with the diagonal component of GF.

\subsection{Off-diagonal excitation and QPE}

When the Gibbs state is input to $\mathcal{C}_{\mathrm{GF-QPE}}$ involving $\mathcal{C}_{m m'}$ for the pair of $m$th and $m'$th WBOs,
the probability that $e \sigma \ (\sigma = +, -)$ excitation occurs and an excitation energy $\varepsilon$ in binary representation is observed is calculated,
from Eq.~(\ref{avr_spectra_using_QPE:circ_GF_QPE_for_energy_eigenstate}),
as
\begin{align}
    \mathbb{P}^{(m m')}_{e \sigma} (\varepsilon)
    &=
        \frac{1}{\mathcal{Z}}
        \sum_{\lambda_0, \lambda}
        e^{-\beta E_{\lambda_0} }
        | \langle \Psi_\lambda | a_{m m'}^{\sigma \dagger} | \Psi_{\lambda_0 } \rangle |^2
        \delta_{E_{\lambda} - E_{\lambda_0}, \varepsilon}
    \nonumber \\
    &=
            \frac{
                S_{m m}^{(e)} (\varepsilon)
                +
                S_{m' m'}^{(e)} (\varepsilon)
                }{4}
            +
            \sigma
            \frac{
                e^{i \pi/4} S_{m m'}^{(e)} (\varepsilon)
                +
                e^{-i \pi/4} S_{m' m}^{(e)} (\varepsilon)
            }{4}
        ,
    \label{avr_spectra_using_QPE:prob_distr_of_exc_energy_for_off_diag_el_QPE}
\end{align}
where we used the definition of the spectral matrix in
Eq.~(\ref{avr_spectra_using_QPE:def_spectral_mat})
to get the second equality.
Similarly, the probability for $h \sigma$ excitation is calculated as
\begin{align}
    \mathbb{P}^{(m m')}_{h \sigma} (\varepsilon)
    &=
        \frac{1}{\mathcal{Z}}
        \sum_{\lambda_0, \lambda}
        e^{-\beta E_{\lambda_0} }
        | \langle \Psi_\lambda | a_{m m'}^\sigma | \Psi_{\lambda_0 } \rangle |^2
        \delta_{E_{\lambda} - E_{\lambda_0}, \varepsilon}
    \nonumber \\
    &=
        \frac{
            S_{m m}^{(h)} (\varepsilon)
            +
            S_{m' m'}^{(h)} (\varepsilon)
            }{4}
        +
        \sigma
        \frac{
            e^{i \pi/4} S_{m m'}^{(h)} (\varepsilon)
            +
            e^{-i \pi/4} S_{m' m}^{(h)} (\varepsilon)
        }{4}
        .
\label{avr_spectra_using_QPE:prob_distr_of_exc_energy_for_off_diag_hole_QPE}
\end{align}
We can solve 
Eqs.~(\ref{avr_spectra_using_QPE:prob_distr_of_exc_energy_for_off_diag_el_QPE})
and
(\ref{avr_spectra_using_QPE:prob_distr_of_exc_energy_for_off_diag_hole_QPE})
with respect to $S_{m m'}^{(\xi)} (\varepsilon)$ to obtain
\begin{align}
    S_{m m'}^{(\xi)} (\varepsilon)
    =
        e^{-i \pi/4}
        \left(
            \mathbb{P}^{(m m')}_{\xi +} (\varepsilon)
            -
            \mathbb{P}^{(m m')}_{\xi -} (\varepsilon)
        \right)
        +
        e^{i \pi/4}
        \left(
            \mathbb{P}^{(m' m)}_{\xi +} (\varepsilon)
            -
            \mathbb{P}^{(m' m)}_{\xi -} (\varepsilon)
        \right)
        \
        (\xi = e, h)
        .
    \label{avr_spectra_using_QPE:prob_distr_with_exc_energy_for_off_diag_QPE}
\end{align}
This equation relates the probability distribution of measurements in $\mathcal{C}_{\mathrm{GF-QPE}}$ with the off-diagonal component of GF.
Although the off-diagonal component $\gamma_{m m'}$ of the one-electron matrix element can be obtained only from $\mathcal{C}_{m m'},$
we have to use it and $\mathcal{C}_{m' m}$ to obtain $S_{m m'}^{(\xi)} (\varepsilon).$

Given Eqs.~(\ref{avr_spectra_using_QPE:prob_distr_with_exc_energy_for_diag_QPE})
and (\ref{avr_spectra_using_QPE:prob_distr_with_exc_energy_for_off_diag_QPE}),
we find that
QPE sampling based on $\mathcal{C}_{\mathrm{GF-QPE}}$ circuits involving the diagonal and off-diagonal excitation circuits
allow us to obtain the GF at a finite temperature without knowing the initial and final states of excitation at each measurement.

\section{Circuit weights expressed in one-electron matrix elements}
\label{sec:Circuit_weights_for_gamma_meas}

For the definitions of circuit weights in 
Eqs.~(\ref{avr_spectra_using_QPE:def_circ_weight_diag}) and
(\ref{avr_spectra_using_QPE:def_circ_weight_off_diag}) for the cost functions,
we can relate those with the measured one-electron matrix elements $\gamma_{\mathrm{meas}}.$

\subsection{Diagonal excitation}

Recalling the relation in
Eq.~(\ref{avr_spectra_using_QPE_ruin:one_el_mat_diag_from_meas})
for the measured diagonal matrix elements and the number of shots,
the numbers of observed electron and hole excitations are
$M_e^{(m)} = M_{\mathrm{d}} (1 - \gamma_{\mathrm{meas}, mm})$
and
$M_h^{(m)} = M_{\mathrm{d}} \gamma_{\mathrm{meas}, mm},$
respectively, for each $m.$
The summations of them over the WBOs are
$
\sum_{m'} M_e^{(m')} 
=
M_{\mathrm{d}} (n_{\mathrm{sorb}} - \mathrm{tr} \gamma_{\mathrm{meas} })
$
and
$\sum_{m'} M_h^{(m')} = M_{\mathrm{d}} \mathrm{tr} \gamma_{\mathrm{meas}},$
where $\mathrm{tr}$ indicates the trace of a matrix.
The weight of electron excitations on the diagonal circuit $\mathcal{C}_m$ is thus written as,
from Eq.~(\ref{avr_spectra_using_QPE:def_circ_weight_diag}),
\begin{align}
    w_e^{(m)}
    =
        \frac{1}{n_{\mathrm{sorb}}}
        \frac{1 - \gamma_{\mathrm{meas}, mm} }{n_{\mathrm{sorb}} - \mathrm{tr} \gamma_{\mathrm{meas}} }
    .
\end{align}
That for hole excitations is written similarly as
\begin{align}
    w_h^{(m)}
    =
        \frac{1}{n_{\mathrm{sorb}}}
        \frac{\gamma_{\mathrm{meas}, m m} }{\mathrm{tr} \gamma_{\mathrm{meas}} }
    .
\end{align}

\subsection{Off-diagonal excitation}

For the off-diagonal excitation circuit,
the probability for observing an electron excitation is,
from Eq.~(\ref{avr_spectra_using_QPE:prob_distr_for_off_diag_excitation}),
$
\mathbb{P}^{(m m')}_e
=    
\mathbb{P}^{(m m')}_{e+} + \mathbb{P}^{(m m')}_{e-}
=
1 - (\gamma_{m m} + \gamma_{m' m'} )/2
$
for the pair of $m$th and $m'$th WBOs,
while that for a hole excitation is
$\mathbb{P}^{(m m')}_h = 1 - \mathbb{P}^{(m m')}_e.$
The numbers of observed electron and hole excitations are thus
$
M_{e+}^{(m m')} + M_{e-}^{(m m')}
=
M_{\mathrm{od}} ( 1 - (\gamma_{\mathrm{meas}, m m} + \gamma_{\mathrm{meas}, m' m'} )/2 )
$
and
$
M_{h+}^{(m m')} + M_{h-}^{(m m')}
=
M_{\mathrm{od}} (\gamma_{\mathrm{meas}, m m} + \gamma_{\mathrm{meas}, m' m'} )/2
,
$
respectively.
The weight of electron excitations on the off-diagonal circuit $\mathcal{C}_{m m'}$ is thus written as,
from Eq.~(\ref{avr_spectra_using_QPE:def_circ_weight_off_diag}),
\begin{align}
    w_e^{(m m')}
    =
        \frac{1}{n_{\mathrm{sorb}} }
        \frac{1 - (\gamma_{\mathrm{meas}, mm} + \gamma_{\mathrm{meas}, m' m'})/2 }{n_{\mathrm{sorb}} - \mathrm{tr} \gamma_{\mathrm{meas}} }
    ,
\end{align}
while that for hole excitations is written as
\begin{align}
    w_h^{(m m')}
    =
        \frac{1}{n_{\mathrm{sorb}} }
        \frac{\gamma_{\mathrm{meas}, mm} + \gamma_{\mathrm{meas}, m' m'} }{2 \mathrm{tr} \gamma_{\mathrm{meas} }}
    .
\end{align}

\section{Natural-orbital representation}
\label{sec:natural_orb_repr}

\subsection{Definition}

The one-electron states
$\{ | \psi_\nu \rangle \}_{\nu = 0}^{n_{\mathrm{sorb}} - 1}$
that diagonalize the one-electron matrix element $\gamma,$
defined in Eq.~(\ref{avr_spectra_using_QPE:def_one_electron_mat_el}),
are called the natural orbitals (NOs) \cite{bib:4643}.
The eigenvalues $\{ n_\nu \}_\nu$ of $\gamma,$ that fall between $0$ and $1,$
are the occupancies of the corresponding NOs.
Specifically, the column vector $\boldsymbol{c}^{(\nu)}$ for the $\nu$th eigenstate satisfies
$\gamma \boldsymbol{c}^{(\nu)} = n_\nu \boldsymbol{c}^{(\nu)}.$
Those vectors form a orthonormalized system since $\gamma$ is Hermitian.
The creation operator of an electron at the $\nu$th NO is given as a linear combination of those for the WBOs:
\begin{align}
    \widetilde{a}_\nu^\dagger
    &=
        \sum_m
            c_m^{(\nu) *}
            a_m^\dagger
    .
    \label{avr_spectra_using_QPE:transform_btwn_MOs_and_NOs_cr}
\end{align}
The annihilation operator for the NO is similarly given by
\begin{align}
\begin{split}
    \widetilde{a}_\nu
    &=
        \sum_m
            c_m^{(\nu)}
            a_m
        .
    \label{avr_spectra_using_QPE:transform_btwn_MOs_and_NOs_an}
\end{split}
\end{align}
They are easily confirmed to satisfy the fermionic anti-commutation relations,
$
\{ \widetilde{a}_\nu, \widetilde{a}_{\nu'}^\dagger \}
=
\delta_{\nu \nu'}
.
$

\subsection{Transition amplitudes}

We define the transition amplitudes for the $m$th WBO as
\begin{align}
\begin{split}
    b^{(\lambda_0 \to \lambda, e)}_m
    &\equiv
        \sqrt{\frac{e^{-\beta E_{\lambda_0} }}{\mathcal{Z}}}
        \langle \Psi_\lambda | a_m^\dagger | \Psi_{\lambda_0 } \rangle
    ,
    \\
    b^{(\lambda_0 \to \lambda, h)}_m
    &\equiv
        \sqrt{\frac{e^{-\beta E_{\lambda_0} }}{\mathcal{Z}}}
        \langle \Psi_\lambda | a_m | \Psi_{\lambda_0 } \rangle
    .
    \label{avr_spectra_using_QPE:def_transition_vec}
\end{split}
\end{align}
We see $b^{(\lambda_0 \to \lambda, e)}_m$ as the $(\lambda_0, \lambda)$th component of a complex vector $\boldsymbol{b}^{(e)}_m.$
We can derive the following sum rule for those electron transition vectors:
\begin{align}
        \boldsymbol{b}^{(e) *}_m
        \cdot
        \boldsymbol{b}^{(e)}_{m'}
    &=
        \langle
            a_m a_{m'}^\dagger
        \rangle
    =
        \delta_{m m'} - \gamma_{m' m}
    ,
    \label{avr_spectra_using_QPE:sum_trans_vec_el_and_1e_mat}
\end{align}
where we used the fact that
$\sum_\lambda | \Psi_\lambda \rangle \langle \Psi_\lambda |$ is the identity.
As for the hole transition vectors $\boldsymbol{b}^{(h)}_m,$
the following sum rule can be derived:
\begin{align}
        \boldsymbol{b}_{m'}^{(h) *}
        \cdot
        \boldsymbol{b}_m^{(h)}
    &=
        \langle a_{m'}^\dagger a_m \rangle
    =
        \gamma_{m' m}
        .
    \label{avr_spectra_using_QPE:sum_trans_vec_hole_and_1e_mat}
\end{align}

Similarly to Eq.~(\ref{avr_spectra_using_QPE:def_transition_vec}),
we define the renormalized transition amplitudes for the $\nu$th NO as
\begin{align}
\begin{split}
    \widetilde{b}_{\nu}^{(\lambda_0 \to \lambda, e)}
    &\equiv
        \frac{1}{\sqrt{1 - n_\nu}}
        \sqrt{\frac{e^{-\beta E_{\lambda_0} }}{\mathcal{Z}}}
        \langle \Psi_\lambda | \widetilde{a}_\nu^\dagger | \Psi_{\lambda_0 } \rangle
    ,
    \\
    \widetilde{b}_{\nu}^{(\lambda_0 \to \lambda, h)}
    &\equiv
        \frac{1}{\sqrt{n_\nu}}
        \sqrt{\frac{e^{-\beta E_{\lambda_0} }}{\mathcal{Z}}}
        \langle \Psi_\lambda | \widetilde{a}_\nu | \Psi_{\lambda_0 } \rangle
    .
    \label{avr_spectra_using_QPE:def_transition_vec_for_NO}
\end{split}
\end{align}
By using
Eqs.~(\ref{avr_spectra_using_QPE:transform_btwn_MOs_and_NOs_cr}),
(\ref{avr_spectra_using_QPE:transform_btwn_MOs_and_NOs_an}),
(\ref{avr_spectra_using_QPE:sum_trans_vec_el_and_1e_mat}),
and
(\ref{avr_spectra_using_QPE:sum_trans_vec_hole_and_1e_mat}),
we can obtain the sum rules of the transition vectors for the NOs as
\begin{align}
        \widetilde{\boldsymbol{b}}_{\nu}^{(\xi) *}
        \cdot
        \widetilde{\boldsymbol{b}}_{\nu'}^{(\xi)}
    &=
        \delta_{\nu \nu'}
    \label{avr_spectra_using_QPE:sum_nat_orb_trans_vec}
\end{align}
for each of $\xi = e, h.$
The sum rules in 
Eq.~(\ref{avr_spectra_using_QPE:sum_nat_orb_trans_vec})
indicate that those vectors are orthonormalized with respect to the NO indices,
in contrast to those for the WBOs.
This feature motivates us to work with the NO representation in the present study since the hyperspherical Householder parametrization in
Appendix \ref{sec:Householder_for_orthogonal_vectors}
is directly applicable to the trial transition amplitudes.

\subsection{Green's function}

The partial GF in the WBO representation is given by Eq.~(\ref{G_imag_freq_sum_partial_G}).
We can obtain that in the NO representation.  
Specifically,
the electron part in the NO representation is
\begin{align}
    \widetilde{G}_{\nu \nu'}^{(e) } (z)
    &=
        \sqrt{1 - n_\nu}
        \sqrt{1 - n_{\nu'}}
        \sum_{\lambda_0, \lambda}
        \frac{
            \widetilde{b}_{\nu}^{(\lambda_0 \to \lambda, e) *}
            \widetilde{b}_{\nu'}^{(\lambda_0 \to \lambda, e)}
        }{ z - ( E_{\lambda} - E_{\lambda_0} )}
    ,
    \label{avr_spectra_using_QPE:gf_in_nat_orb_repr_e}
\end{align}
while the hole part is
\begin{align}
    \widetilde{G}_{\nu \nu'}^{(h) } (z)
    &=
        \sqrt{n_\nu}
        \sqrt{n_{\nu'} }
        \sum_{\lambda_0, \lambda}
        \frac{
            \widetilde{b}_{\nu'}^{(\lambda_0 \to \lambda, h) *}
            \widetilde{b}_{\nu}^{(\lambda_0 \to \lambda, h)}
        }{ z - ( E_{\lambda_0} - E_{\lambda} )}
    .
    \label{avr_spectra_using_QPE:gf_in_nat_orb_repr_h}
\end{align}
Since the transition amplitudes in the WBO and NO representations are related to each other via the linear transformations in 
Eqs.~(\ref{avr_spectra_using_QPE:transform_btwn_MOs_and_NOs_cr}) and
(\ref{avr_spectra_using_QPE:transform_btwn_MOs_and_NOs_an}),
the GF for $\xi$ excitation $(\xi = e, h)$ in the NO representation is calculated as 
\begin{align}
    \widetilde{G}_{\nu \nu'}^{(\xi) } (z)
    =
        \sum_{m, m'}
            c^{(\nu)}_m
            G_{m m'}^{(\xi)} (z)
            c^{(\nu') *}_{m'}
        .
\end{align}
We define the unitary matrix $U_{\mathrm{nat}},$
where the row vector $\boldsymbol{c}^{(\nu) \mathrm{T}}$ is the $\nu$th row.
We can rewrite the equation above as
\begin{align}
    \widetilde{G}^{(\xi) } (z)
    =
        U_{\mathrm{nat}}
        G^{(\xi) } (z)
        U_{\mathrm{nat}}^\dagger
        .
    \label{avr_spectra_using_QPE:relation_of_gf_btwn_MOs_and_NOs}
\end{align}

\section{Green's function from quadratic fictitious density of states}
\label{sec:gf_from_fict_dos}

Here we consider a quadratic fictitious density of states (DOS)
\begin{align}
    \rho_{\mathrm{quad}} (E; \Delta E)
    \equiv
    \begin{cases}
        -\frac{6}{\Delta E^3} E^2 + \frac{3}{2 \Delta E}
        &
        | E | \leq \Delta E/2
        \\
        0
        &
        \mathrm{otherwise}
    \end{cases}
\end{align}
centered at the origin of the energy axis.
$\Delta E$ is its spectral width.
This DOS is normalized to be unity as
$\int_{-\infty}^\infty d E \rho_{\mathrm{quad}} (E; \Delta E) = 1.$
The contribution from the DOS located around an excitation energy $E_0$ to the one-particle GF is calculated as
\begin{align}
    g_{\mathrm{quad}} (z; E_0, \Delta E)
    &\equiv
        \int_{-\infty}^\infty
        d E
            \frac{\rho_{\mathrm{quad}} (E - E_0; \Delta E)}{z - E}
    \nonumber \\
    &=
        \frac{1}{\Delta E}
        \int_{-1/2}^{1/2}
            \frac{dx}{\widetilde{z} - x}
            \left( -6 x^2 + \frac{3}{2} \right)
        \
        \left(
            \widetilde{z} \equiv (z - E_0)/\Delta E
        \right)
    \label{gf_from_fict_dos:gf_from_quad_fict_dos}
\end{align}
for a complex frequency $z.$
There exists an integral formula
\begin{align}
    \int
    dx
    \frac{c_0 - x^2}{\alpha - x}
    &=
        \alpha x
        + \frac{x^2}{2}
        +
        i (-c_0 + \alpha^2)
        \arctan \frac{x - \mathrm{Re} \alpha}{\mathrm{Im} \alpha}
        +
        \frac{-c_0 + \alpha^2}{2}
        \ln
        \left(
            1 + \frac{(x - \mathrm{Re} \alpha)^2}{(\mathrm{Im} \alpha)^2}
        \right)
        +
        \mathrm{const.}
\end{align}
for complex $c_0$ and $\alpha$ with $-\pi/2 \leq \arctan \leq \pi/2.$
We can thus calculate Eq.~(\ref{gf_from_fict_dos:gf_from_quad_fict_dos}) analytically as
\begin{gather}
    g_{\mathrm{quad}} (z; E_0, \Delta E)
    =
        \frac{6}{\Delta E}
        \widetilde{z}
        +
        g_{2+} (z) - g_{2-} (z)
    ,
\end{gather}
where we defined      
\begin{align}
    g_{2 \pm} (z)
    \equiv
        \frac{6}{\Delta E}
        i
        \left( \widetilde{z}^2 - \frac{1}{4} \right)
        \arctan
        \frac{\mathrm{Re} \widetilde{z} \pm 1/2}{\mathrm{Im} \widetilde{z} } 
        -
        \frac{6}{\Delta E}
        \frac{1}{2}
        \left( \widetilde{z}^2 - \frac{1}{4} \right)
        \ln
        \left(
            1
            +
            \frac{(\mathrm{Re} \widetilde{z} \pm 1/2)^2}{(\mathrm{Im} \widetilde{z})^2 } 
        \right)
    .
\end{align}
Figure \ref{fig:gf_fict_dos} shows the shape of $g_{\mathrm{quad}}$ as a function of a complex frequency having a small imaginary part.
We can see $\mathrm{Im} g_{\mathrm{quad}}$ exhibit the blunt shape coming from the finite spectral width of the fictitious DOS.

\begin{figure}
\begin{center}
\includegraphics[width=6cm]{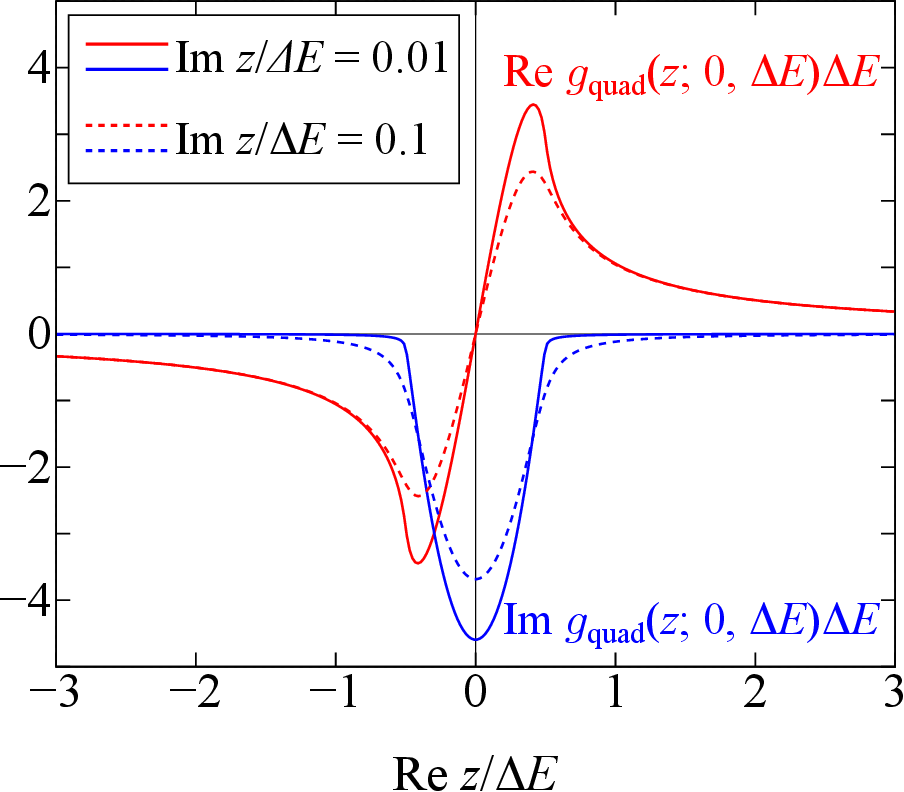}
\end{center}
\caption{
Green's function coming from the fictitious DOS $\rho_{\mathrm{quad}}$ centered at the origin with a width $\Delta E.$
}
\label{fig:gf_fict_dos}
\end{figure}

\section{Hyperspherical Householder parametrization for orthonormalized vectors}
\label{sec:Householder_for_orthogonal_vectors}

The Householder parametrization \cite{bib:7129,bib:7123} is a useful technique for parametrizing an arbitrary number of orthonormalized vectors in arbitrary-dimensional Euclidean space. 

We first review the generic Householder parametrization.
Let us assume that we want to parametrize $n$ orthonormalized vectors in $\mathbb{R}^m \ (n \leq m).$
For each of $k = 0, \dots, n - 1,$ we construct a unit vector
\begin{align}
    \boldsymbol{u}^{(k)}
    \equiv
        \begin{pmatrix}
            \boldsymbol{0}_k \\
            \boldsymbol{e}^{(k)}        
        \end{pmatrix}
    \in
    \mathbb{R}^m
    \label{Householder_for_orthogonal_vectors:def_u}
\end{align}
from any unit vector $\boldsymbol{e}^{(k)} \in \mathbb{R}^{m - k}.$
We then define the reflection matrix
\begin{align}
    H^{(k)}
    \equiv
        I_m
        -
        2
        \boldsymbol{u}^{(k)}
        \boldsymbol{u}^{(k) \mathrm{T}}
    =
        \begin{pmatrix}
            I_k & O_{k \times (m - k)} \\
            O_{(m - k) \times k} & I_{m - k} - 2 \boldsymbol{e}^{(k)} \boldsymbol{e}^{(k) \mathrm{T} }
        \end{pmatrix}
        \in \mathbb{R}^{m \times m}
        ,
\end{align}
where $I_\ell$ is the identity in $\mathbb{R}^\ell$ and $O_{\ell \times \ell'}$ is the zero matrix.
$H^{(k)}$ is easily confirmed to be symmetric and orthogonal.
From $I_{m \times n},$ whose $n$ diagonal components are $1,$
and the reflection matrices,
we construct
\begin{align}
    Q
    \equiv
        H^{(0)}
        H^{(1)}
        \cdots
        H^{(n - 1)}
        I_{m \times n}
    \in \mathbb{R}^{m \times n}
    .
\end{align}
This matrix is orthogonal, that is,
$Q^{\mathrm{T}} Q = I_n$ and the $n$ column vectors contained in $Q$ form an orthonormal system for $\mathbb{R}^m.$
The specification of $Q$ via the unit vectors $\{ \boldsymbol{e}^{(k)} \}_{k = 0}^{n - 1}$ in this way is the Householder parametrization,
which allows us to obtain orthonormalized vectors without imposing explicit constraints on the unit vectors.
Since the number of independent components in $\boldsymbol{e}^{(k)}$ is $m - k - 1$ due to the normalization condition,
the degree of freedom for specifying $Q$ is
\begin{align}
    \sum_{k = 0}^{n - 1} (m - k -1) 
    =
        m n - \frac{n (n + 1)}{2}    
    .
    \label{Householder_for_orthogonal_vectors:num_of_params}
\end{align}

In the present study,
we adopt hyperspherical parametrization as a special case of the technique described just above.
Specifically, for each $k,$ we introduce $m - k - 1$ angle parameters 
$\varphi^{(k)}_{ 0}, \dots, \varphi^{(k)}_{m - k - 2}$ to locate a point
on a $(m - k - 1)$-dimensional sphere in $\mathbb{R}^{m - k}$ as
\begin{align}
\begin{split}
    e^{(k)}_0
    &=
        \cos \varphi^{(k)}_{ 0}
    ,
    \\
    e^{(k)}_1
    &=
        \sin \varphi^{(k)}_{ 0} \cos \varphi^{(k)}_{ 1}
    ,
    \\
    e^{(k)}_2
    &=
        \sin \varphi^{(k)}_{ 0} \sin \varphi^{(k)}_{ 1} \cos \varphi^{(k)}_{ 2}
    ,
    \\
    &\vdots
    \\
    e^{(k)}_{m - k - 2}
    &=
        \sin \varphi^{(k)}_{ 0} \sin \varphi^{(k)}_{ 1} \sin \varphi^{(k)}_{ 2}
        \cdots
        \cos \varphi^{(k)}_{ m - k - 2}
    ,
    \\
    e^{(k)}_{m - k - 1}
    &=
        \sin \varphi^{(k)}_{ 0} \sin \varphi^{(k)}_{ 1} \sin \varphi^{(k)}_{ 2}
        \cdots
        \sin \varphi^{(k)}_{ m - k - 2}
    .
\end{split}
\end{align}
We use the unit vector $\boldsymbol{e}^{(k)}$ defined in this way for
Eq.~(\ref{Householder_for_orthogonal_vectors:def_u}).
This parametrization allows us to obtain all the possible orthonormalized vectors with the periodic parameter ranges.

\section{Modeled probability distributions}
\label{sec:modeled_prob_distrs}

\subsection{Reconstructed GF in WBO representation}

By employing the relation between the true GFs in the WBO and NO representations in
Eq.~(\ref{avr_spectra_using_QPE:relation_of_gf_btwn_MOs_and_NOs}),
we can get the reconstructed GF
$G_{\mathrm{rec} } (z; \Lambda_e)$
in the WBO representation from
Eqs.~(\ref{avr_spectra_using_QPE:gf_rec_in_NO_using_DOS_e}) and
(\ref{avr_spectra_using_QPE:gf_rec_in_NO_using_DOS_h}) as
\begin{align}
\begin{split}
    G_{\mathrm{rec} }^{(e)} (z; \Lambda_e)
    &=
        \sum_{\ell = 0}^{n_{\mathrm{ch}} - 1 }
        \int_{-\infty}^\infty
        dE
        \rho_{e \ell} (E - \varepsilon_{e \ell})
        \frac{W_{\ell}^{(e)} }{z - E }
    ,
    \\
    G_{\mathrm{rec} }^{(h)} (z; \Lambda_h)
    &=
        \sum_{\ell = 0}^{n_{\mathrm{ch}} - 1 }
        \int_{-\infty}^\infty
        dE
        \rho_{h \ell} (E - \varepsilon_{h \ell})
        \frac{W_{\ell}^{(h)} }{z + E }
    ,
    \label{avr_spectra_using_QPE:gf_rec_in_MO_using_DOS}
\end{split}
\end{align}
where we defined the matrices $W_{\ell}^{(e)}$ and $W_{\ell}^{(h)}$ via
\begin{align}
\begin{split}
    W_{\ell, m m'}^{(e)}
    &\equiv
        \sum_{\nu, \nu'}
        (U_{\mathrm{nat}}^\dagger)_{m \nu}
        \sqrt{1 - n_{\mathrm{meas}, \nu}}
        \sqrt{1 - n_{\mathrm{meas}, \nu'}}
            v^{(\nu)}_{e \ell}
            v^{(\nu')}_{e \ell}
        U_{\mathrm{nat} \nu' m'}
    ,
    \\
    W_{\ell, m m'}^{(h)}
    &\equiv
        \sum_{\nu, \nu'}
        (U_{\mathrm{nat}}^\dagger)_{m \nu}
        \sqrt{n_{\mathrm{meas}, \nu} }
        \sqrt{n_{\mathrm{meas}, \nu'} }
            v^{(\nu)}_{h \ell}
            v^{(\nu')}_{h \ell}
        U_{\mathrm{nat} \nu' m'}
    .
\end{split}
\end{align}
Recalling the relation between the true GF and spectral matrices in
Eqs.~(\ref{avr_spectra_using_QPE:GF_el_using_kronecker_delta}) and
(\ref{avr_spectra_using_QPE:GF_hole_using_kronecker_delta}),
we deduce the reconstructed spectral matrices in the WBO representation
from Eq.~(\ref{avr_spectra_using_QPE:gf_rec_in_MO_using_DOS}) as
\begin{align}
    S_{\mathrm{rec}, m m' }^{(\xi)} (E)
    &=
        \sum_{\ell = 0}^{n_{\mathrm{ch}} - 1 }
        \rho_{\xi \ell} (E - \varepsilon_{\xi \ell})
        W_{\ell, m m'}^{(\xi)}
    .
    \label{avr_spectra_using_QPE:spec_mat_rec_in_MO}
\end{align}

We can derive the sum rule satisfied by the $W_{\ell}^{(e)}$ matrices that will be useful for debugging in numerical implementation.
Specifically, we sum up their traces over the excitation channels to obtain
\begin{align}
    \sum_{\ell = 0}^{n_{\mathrm{ch}} - 1}
    \mathrm{tr}
        W^{(e)}_{\ell}
    &=
        \sum_{\ell = 0}^{n_{\mathrm{ch}} - 1}
        \sum_{\nu = 0}^{n_{\mathrm{sorb}} - 1}
        (1 - n_{\mathrm{meas}, \nu})
        v^{(\nu)}_{e \ell}
        v^{(\nu)}_{e \ell}
    \nonumber \\
    &=
        n_{\mathrm{sorb}}
        -
        \mathrm{tr} \gamma_{\mathrm{meas} }
    ,
\end{align}
where we used the cyclic nature of a trace to get the first equality and the orthonormality between the fictitious transition vectors in 
Eq.~(\ref{avr_spectra_using_QPE:orthonormality_btwn_fict_trans_vecs})
to get the second equality.
We can also derive the sum rule satisfied by the $W_{\ell}^{(h)}$ matrices similarly as
\begin{align}
    \sum_{\ell = 0}^{n_{\mathrm{ch}} - 1}
    \mathrm{tr}
        W^{(h)}_{\ell}
    =
        \mathrm{tr} \gamma_{\mathrm{meas} }
        .
\end{align}

\subsection{Diagonal excitation and QPE}

The true probability distribution for the diagonal excitation circuit $\mathcal{C}_m$ and the spectral matrix are related via
Eq.~(\ref{avr_spectra_using_QPE:prob_distr_with_exc_energy_for_diag_QPE}),
while the reconstructed spectral matrix is given by 
Eq.~(\ref{avr_spectra_using_QPE:spec_mat_rec_in_MO}).
It is thus reasonable to model the probability distribution
for $\xi$ excitations and subsequent QPE measurements for a fictitious input state as
\begin{align}
    \mathbb{P}^{(p, m)}_{\xi j}
    ( \Lambda_\xi )
    \equiv
        \int_{-\infty}^\infty
        d E \,
        S_{\mathrm{rec}, m m }^{(\xi)} (E)
        \mathbb{P}^{(p)}_j (E)
    =
        \sum_{\ell = 0}^{n_{\mathrm{ch}} - 1}
            W^{(\xi)}_{\ell, m m}
            \mathbb{P}_{\xi \ell j}^{(p)}
    ,
    \label{avr_spectra_using_QPE:prob_distr_for_diag_from_trial_values}
\end{align}
where $p$ indicates the setting of QPE,
$\xi = e, h$ indicates the excitation, and
$j = 0, \dots, N_{\mathrm{val}} - 1$ is for the grid points represented by the ancillae of QPE. 
\begin{gather}
    \mathbb{P}^{(p)}_j (E)
    \equiv
        \left|
            \frac{1}{N_{\mathrm{val}}}
            \sum_{j' = 0}^{N_{\mathrm{val}} - 1}
            \exp
            \frac{i 2 \pi j' ((E - E_{\mathrm{orig}}^{(p)}) t_0 - j)}{N_{\mathrm{val}} }
        \right|^2
\end{gather}
is the probability for observing $j$ as a bit string when the excitation energy is $E - E_{\mathrm{orig}}^{(p)}$ \cite{Nielsen_and_Chuang}.
\begin{align}
    \mathbb{P}_{\xi \ell j}^{(p)}
    \equiv
        \int_{-\infty}^\infty
        dE
        \rho_{\xi \ell} (E - \varepsilon_{\xi \ell})
        \mathbb{P}^{(p)}_j (E)    
\end{align}
is the probability that the observed ancillae represent $j$ as a bit string when the $\ell$th $\xi$ excitation channel is invoked.

The probability distribution defined in 
Eq.~(\ref{avr_spectra_using_QPE:prob_distr_for_diag_from_trial_values})
can be confirmed to satisfy the sum rule
\begin{align}
    \sum_{\xi = e, h}
    \sum_{j = 0}^{N_{\mathrm{val}} - 1}
    \mathbb{P}^{(p, m)}_{\xi j}
    ( \Lambda_\xi )
    =
        1
\end{align}
due to $\sum_j \mathbb{P}^{(p)}_j (E) = 1$ and the normalization conditions of the fictitious DOS.

\subsection{Off-diagonal excitation and QPE}

The ture probability distribution for the off-diagonal excitation circuit $\mathcal{C}_{m m'}$ and the spectral matrix are related via
Eqs.~(\ref{avr_spectra_using_QPE:prob_distr_of_exc_energy_for_off_diag_el_QPE}) and (\ref{avr_spectra_using_QPE:prob_distr_of_exc_energy_for_off_diag_hole_QPE}).
We therefore model the probability distribution
for $\xi \sigma$ excitations and subsequent QPE measurements for a fictitious input state as
\begin{align}
    \mathbb{P}^{(p, m m')}_{\xi \sigma j}
    ( \Lambda_\xi )
    &\equiv
        \int_{-\infty}^\infty
        d E \,
    \left(
        \frac{
            S_{\mathrm{rec}, m m}^{(\xi)} (E)
            +
            S_{\mathrm{rec}, m' m'}^{(\xi)} (E)
            }{4}
        +
        \sigma
        \frac{1}{2}
        \mathrm{Re}
        \left(
            e^{i \pi/4}
            S_{\mathrm{rec}, m m'}^{(\xi)} (E)
        \right)
    \right)
        \mathbb{P}^{(p)}_j (E)
    \nonumber \\
    &=
        \sum_{\ell = 0}^{n_{\mathrm{ch}} - 1}
    \left(
        \frac{
            W^{(\xi)}_{\ell, m m}
            +
            W^{(\xi)}_{\ell, m' m'}
        }{4}
        +
        \sigma
        \frac{1}{2}
        \mathrm{Re}
        \left(
            e^{i \pi/4}
            W^{(\xi)}_{\ell, m m'}
        \right)
    \right)
        \mathbb{P}_{\xi \ell j}^{(p)}
    \label{avr_spectra_using_QPE:prob_distr_for_off_diag_from_trial_values}
\end{align}
for $\sigma = +, -.$
We used the expression for the reconstructed spectral matrix in
Eq.~(\ref{avr_spectra_using_QPE:spec_mat_rec_in_MO}) to get the second equality above.
This distribution can be confirmed to satisfy the sum rule
\begin{align}
    \sum_{\xi = e,h}
    \sum_{\sigma = +,-}
    \sum_{j = 0}^{N_{\mathrm{val}} - 1}  
        \mathbb{P}^{(p, m m')}_{\xi \sigma j}
        ( \Lambda_\xi )
    =
        1
        .
\end{align}

\end{widetext}

\bibliography{ref}

\end{document}